\documentclass[english]{article}
\usepackage[utf8]{inputenc}
\DeclareUnicodeCharacter{B0}{\textdegree}
\DeclareUnicodeCharacter{FB01}{fi}
\usepackage{amsmath,amssymb,amsfonts}
\usepackage[T1]{fontenc}
\usepackage{color}
\usepackage{amsmath,comment}
\usepackage{graphicx}
\usepackage{listings}
\usepackage[ruled,vlined,linesnumbered]{algorithm2e}
\usepackage[section]{placeins}

\definecolor{codegreen}{rgb}{0,0.6,0}
\definecolor{codegray}{rgb}{0.5,0.5,0.5}
\definecolor{codepurple}{rgb}{0.58,0,0.82}
\definecolor{backcolour}{rgb}{0.95,0.95,0.92}
 
\lstdefinestyle{sourcecode}{
    backgroundcolor=\color{backcolour},   
    commentstyle=\color{codegreen},
    numberstyle=\tiny\color{codegray},
    stringstyle=\color{codepurple},
    basicstyle=\ttfamily\normalsize,
    breakatwhitespace=false,         
    breaklines=true,                 
    captionpos=b,                    
    keepspaces=true,                 
    numbers=left,                    
    numbersep=5pt,                  
    showspaces=false,                
    showstringspaces=false,
    showtabs=false,                  
    tabsize=2
}
 
\makeatletter
\@ifundefined{date}{}{\date{}}

\usepackage{babel}
\definecolor{MyDarkGreen}{rgb}{0.0,0.4,0.0}

\usepackage{babel}
\usepackage{listings}
\usepackage{babel}
\usepackage{listings}

\usepackage{babel}
\usepackage{listings}

\makeatother

\usepackage{babel}
\usepackage{authblk}
\usepackage{listings}

\begin{document}

\title{\vspace{-4cm}Estimations of Integrated Information Based on Algorithmic Complexity and Dynamic Querying}

\author[1]{Alberto Hern\'andez-Espinosa}
\author[2,3,4,5]{Hector Zenil\thanks{Co-first and corresponding author: hector.zenil [at] algorithmicnaturelab [dot] org}}
\author[2,3,4]{Narsis A. Kiani}
\author[3,6]{Jesper Tegn\'er}
\affil[1]{Depto. de Matem\'aticas, Facultad de Ciencias, UNAM, M\'exico;}
\affil[2]{Algorithmic Dynamics Lab, Karolinska Institute (KI);}
\affil[3]{Unit of Computational Medicine, Center for Molecular Medicine, Department of Medicine Solna, KI, Stockholm, Sweden;}
\affil[4]{Algorithmic Nature Group, Laboratory of Scientific Research (LABORES) for the Natural and Digital Sciences, Paris, France;}
\affil[5]{Oxford Immune Algorithmics, Reading, U.K.;}
\affil[6]{Biological and Environmental Sciences and Engineering Division\\Computer, Electrical and Mathematical Sciences and Engineering\\Division, King Abdullah University of Science and\\ Technology (KAUST), Kingdom of Saudi Arabia.}
\date{}

\maketitle

\vspace{-1cm}
\begin{abstract}
\textcolor{black}{
Information has emerged as a language in its own right, bridging several disciplines that analyze natural phenomena and man-made systems. Integrated information has been introduced as a metric to quantify the amount of information generated by a system beyond the information generated by its elements. Yet, this intriguing notion comes with the price of being prohibitively expensive to calculate, since the calculations require an exponential number of sub-divisions of a system. Here we introduce a novel framework to connect algorithmic randomness and integrated information, and a numerical method for estimating integrated information using a perturbation test rooted in algorithmic information dynamics. This method quantifies the change in program size of a system, when subjected to a perturbation. The intuition behind is that if an object is random then random perturbations have little to no effect to what happens when a shorter program but when an objects has the ability to move in both directions (towards or away from randomness) it will be shown to be better integrated as a measure of sophistication telling apart randomness and simplicity from structure. We show that an object with a high integrated information value is also more compressible, and is therefore more sensitive to perturbations. We find that such a perturbation test quantifying compression sensitivity, provides a system with a means to extract explanations--causal accounts--of its own behaviour. Our technique can reduce the number of calculations to arrive to some bounds or estimations, as the algorithmic perturbation test guides an efficient search for estimating integrated information. Our work sets the stage for a systematic exploration of connections between algorithmic complexity and integrated information at the level of both theory and practice.}
\\


\end{abstract}

\section{Introduction}

\textcolor{black}{The development of techniques to decipher the structure and dynamics of complex systems is a rich inter-disciplinary research area which is not only of fundamental interest but also important in numerous applications. Broadly speaking, dynamical aspects such as stability and state-transitions within such systems have been of major interest in statistical physics, dynamical systems, and computational neuroscience \cite{strogatz,dayan,chandler}. Here, complex systems are defined by a set of non-linear evolution equations. Cellular automata, spin-glass systems, Hopfield networks, and Boolean networks, have for example been used as numerical experimental model systems to investigate the dynamical aspects of complex systems. Due to the complexity of the analysis, notions such as symmetries in the systems, averaging (e.g. mean-field techniques), and separation of time-scales, have all been instrumental in deciphering the core principles at work in such complex systems. In parallel, network science has emerged as a rich interdisciplinary field, essentially analyzing the structure of real networks in different areas of science and in diverse application domains \cite{barabasi}. Examples include social, biological and electrical networks, the web, business networks and the interconnected internet. 
By a structural analysis, which has dominated these investigations, we refer to statistical descriptions of network connectivity. Networks can be described globally, in terms ranging from the degree to which they differ from a random Poisson distribution of links, to their modular organization, including their local properties such as local clustering around nodes, special nodes/links with high degrees of betweenness or serving specific roles in the network, and local motif structures. Such network features can be used to classify and describe similarities and differences between what appear to be different classes of networks across and within different application domains. Finally, due to the rich representational capacity of networks and their usefulness across science, technology, and applications, work in machine learning, in particular graph convolutional networks and embedding techniques, is currently making headway in devising ways to map these non-regular network objects onto a format such that machine learning techniques can be used to analyze their properties~\cite{Bronstein}.}

\textcolor{black}{Now, we may ask if integrated information theory (IIT) is proposed to be of relevance for the analysis of complex networks, we ask how is IIT related to fundamental questions underpinning research and thinking of complex systems? On the one hand, we find a rich body of work dealing with what could be referred to as technical, computational challenges, and application-driven investigations. For example, which global and local properties should be computed and how to do so in an efficient manner. However, at a more fundamental level we find essentially two challenges, which in our view have a bearing on
the core intellectual driving force of complex systems. First: What is the origin of and mechanisms propelling order in complex systems? Secondly, and of major concern for the present paper: Is the whole - in some sense - larger than the sum of its parts? Both questions are vague when formulated in words, as above, but they can readily be technically specified within a model class. The motivation for the second question is that it appears that there are indeed phenomena in nature which cannot easily be explained only with reference to their parts, but seem to require that we adopt a holistic view. Since Anderson's classic 1972 essay, there has been an animated and at times heated discussion of whether there is anything which could be referred to as emergence.\cite{Anderson}}

\textcolor{black}{Tononi and his group have developed a formalism--targeting exactly such a holistic analysis--specifically to quantify the amount of information generated by a system -- defined as a set of interconnected elements--beyond the information generated by the parts (subsets) of the system. Their motivation was that in order to develop a theory of consciousness~\cite{oizumi2014phenomenology}. In that quest, they perceived a necessity to define a measure which could quantify the amount and degree of consciousness, a measure they refer to as $\phi$, which in turn constitutes the core of Integrated Information Theory or IIT. Importantly, in the present work we distinguish between the issue of the relevance of $\phi$ for consciousness versus the technical numerical question of how to calculate $\phi$. Here we address the computation of $\phi$, as it is potentially a means toward a precise formulation for the possible causal relation between a whole and the parts of a system, regardless of its purported relevance to consciousness. To calculate $\phi$, Tononi and collaborators have developed a computational toolbox~\cite{mayner2017pyphia}. Yet, calculating $\phi$ comes with a severe computational cost, as the calculation scales exponentially with the number of elements in the network. Furthermore, the computation requires knowledge of the transition probabilities of the system, which makes computation of anything larger than small systems of order of one magnitude intractable in practice. The calculation of $\phi$ requires a division of the system into smaller subsets, ranging from large pieces down to singletons, 
every division into $k$ pieces can be instantiated in $\dbinom{N}{k}$ different ways. Using this procedure from Tononi, elements that have small causal influences on the activity of other elements can be identified. A system with low $\phi$ is therefore characterized by the fact that changes in subsets of the system do not affect the rest of the system. Such a system is therefore considered to be a non-integrated system. This observation entails a key insight, namely, that if a system is highly integrated among its parts, then the different parts can be related to each other, or more precisely, they can be used to describe other parts of the system. Then the parts are in some sense simple and should be compressible.}

\textcolor{black}{This is the observation and intuition behind our method, which employs a formalized notion of complexity to exploit this insight and thereby allow a more efficient, guided search in the space of algorithmic distances, in contrast to exhaustive computations of the distance between statistical distributions, as currently implemented in IIT. Technically we are therefore not required to perform a full computation of what is referred to as the input-output repertoire (see Methods for technical details). This, in brief, is our motivation for introducing our method, which is based on algorithmic information dynamics~\cite{maininfo,zenilaid,mainbook}. At its core is a causal perturbation analysis and a measure of sophistication connected to algorithmic complexity. Our approach exploits the idea that causal deterministic systems have a simple algorithmic description and thus a simple generating mechanism sufficient to simulate and reproduce complex systemic behaviour. Using this technique we can assess the effect of perturbations, and thereby exploit the fact that, depending on the algorithmic complexity of a system, the perturbation will induce different degrees of change in algorithmic space. In short, a system will be highly integrated only if the removal or perturbation of its parts has a non-linear effect on the generative program producing the system in the first place.}
\textcolor{black}{Interestingly, even Tononi suggested early on that algorithmic complexity could be connected to the computation of integrated information ~\cite{sciencetranslationmedicine}. However, a lossless compression algorithm was used to approximate $\phi$. Here we contribute to the formalization of such a suggestion by using stronger tools, which we have recently developed, to approximate complexity. 
At the core of algorithmic information is the concept of minimal program-size and Kolmogorov-Chaitin complexity~\cite{kolmogorov,chaitin}. Briefly, the Kolmogorov-Chaitin complexity $K(x)$ of an object $x$ is the length of the shortest computer program that produces $x$ and halts. $K(x)$ is uncomputable but can be approximated from above, meaning one can find upper bounds by using compression algorithms, or rather more powerful techniques such as those based on algorithmic probability~\cite{d4,d5,bdm}, given that popular lossless compression algorithms are limited and more closely related to classical Shannon entropy than to $K$ itself~\cite{zkpaper,smalldata,liliana}. One reason for this state of affairs is that, as demonstrated in \cite{shalizi2001computational}, there is a fundamental difference between algorithmic and statistical complexity with respect to how randomness is characterised in opposition to causation. Specifically, algorithmic complexity implies a deterministic description of an object (it defines the algorithmic information content of an individual sequence/object), whereas statistical complexity implies a statistical description (it refers to an ensemble of sequences generated by a certain source). Approaches such as transfer entropy~\cite{transfer}, Granger causality~\cite{granger}, and Partial Information Decomposition~\cite{williams,williams2} that are based on regression, correlation and/or a combination of regression, correlation and intervention but ultimately relying on probability distributions, fall into this category. Hence for better-founded methods and algorithms for estimating algorithmic complexity, we recommend the use of our tools, which are already being used by independent groups working on, for example, biological modelling~\cite{victor}, cognition~\cite{ventresca} and consciousness~\cite{ruffini}. These tools are based on the theory of algorithmic probability, and are not free from challenges and limitations, but they are better connected to the algorithmic side of algorithmic complexity, rather than only to the statistical pattern-matching side that current approaches using popular lossless compression algorithms exploit, making these approaches potentially misleading~\cite{zkpaper}.}
\textcolor{black}{Our procedure, in brief, is as follows. First, we deduce the rules in systems of interest: we apply the perturbation test introduced in \cite{zenilturingtest,maininfo,mainbook} to ascertain the computational capabilities of networks. Next, simple rules are formalized and implemented to simulate the behaviour of these systems. Following this analysis, we perform an automatic procedure, referred to as a meta-perturbation test, which is applied over the behaviour obtained by the aforementioned simple rules, in order to arrive at explanations of such behaviour.} We incorporate the ideas
of an interventionist calculus \textcolor{black}{(c.f. Judea Pearl~\cite{pearl})} and perturbation analysis within what we call Algorithmic Information Dynamics, and \textcolor{black}{we go beyond pattern identification using probability theory, classical statistics, and correlation analysis by developing a model-driven approach that is fed by data. This contrasts with a purely data-driven approach, and is a consequence of the fact that our analysis considers the algorithmic distance between models.}

\section{Basic concepts - Integrated Information Theory (IIT)}


\textcolor{black}{Integrated information theory (IIT) postulates that
consciousness is identical to integrated information and that a system's
capacity for consciousness can be expressed by a quantitative measure
denoted by $\phi$}. \textcolor{black}{Tononi defines integrated information as ``the amount
of information generated by a complex of elements, above and beyond the information generated by its parts'' (Consciousness as integrated
information: a provisional manifesto. Tononi 2008) and states, }\textcolor{black}{\emph{``The integrated information theory}}\textcolor{black}{{} (IIT) of consciousness
claims that, at a fundamental level, consciousness is integrated information'' (Consciousness as integrated information: a provisional
manifesto. Tononi 2008, italics in original).} \textcolor{black}{IIT aims to explain ``relationships between consciousness and the Physical Substrate of Consciousness (PSC), and starts from essential properties of phenomenal experience, and derives the
requirements for the physical substrate of consciousness.''~\cite{tononi2016}}

\textcolor{black}{A first formulation consisted in 5 axioms, which are held to be self-evident within IIT:}
\begin{enumerate}
\item \textcolor{black}{Existence. It is not possible to deny the existence of consciousness, and that consciousness exists is self-evident from within its own perspective.}
\item \textcolor{black}{Composition. Each experience has components.}
\item \textcolor{black}{Information: An experience is specific to certain things, distinct from others.}
\item \textcolor{black}{Integration. Consciousness is irreducible to separate
elements.}
\item \textcolor{black}{Exclusion. Because consciousness specifies certain things, it excludes others; at the same time, it flows with a specific velocity.}
\end{enumerate}
\textcolor{black}{From these axioms IIT formulates five postulates, in essence making causal claims:}
\begin{enumerate}
\item \textcolor{black}{The existence of consciousness implies a system of mechanisms with a particular cause-effect power.}
\item \textcolor{black}{The 
composite nature of consciousness implies that its system's mechanistic elements must have the capacity to combine, and that those combinations have cause-effect power.}
\item \textcolor{black}{Because consciousness is informative, it must specify, or distinguish one experience from another.}
\item \textcolor{black}{Because consciousness is integrated, every one of its element must have the capacity to act as a cause on the rest of the system and to be affected by the rest of the system. If a system can be divided into two parts without affecting its cause-effect structure, it fails to satisfy the requirement of this postulate.}
\item \textcolor{black}{Because of its exclusivity, the various simultaneous subsets of mechanisms in a system have varying cause-effect structures. }
\end{enumerate}
\textcolor{black}{Since the postulates make causal claims, it is paramount to establish a quantitative procedure to assess whether or not such a cause-effect structure is present. Thus, is the system integrated with respect to its internal cause-effect organization? To this end, IIT introduces the parameter $\phi$ as a measure of the amount of
integrated information and hence the level of consciousness of the system. The measure $\phi$ is calculated by dividing the set of elements of a model of the system progressively into subsets in order to identify elements whose (de)activation has few causal consequences upon the (de)activation of other elements. Then, if no logically possible partition of the system results in a loss of connection, the conclusion is that the
system has no positive $\phi$ value. In other words, changes in subparts of a system that do not affect the rest of the system identify a non-integrated
system.}

\subsubsection*{\textcolor{black}{A $\phi$ estimation calculus}}

\textcolor{black}{The integrated information theory defines integrated information ($\phi$) as the effective information of the minimum information partition (MIP) in a system (Tononi, 2004, 2012; Oizumi et al., 2014, 2016a; Tononi et al., 2016). The MIP is also defined as the partition having minimum effective information among all possible partitions.}
\begin{center}
\textcolor{black}{$\phi[X;x]=:\varphi[X;x,MIP(x)]$}
\par\end{center}

\begin{center}
\textcolor{black}{$MIP(x)=:argmin{\varphi(X;x,P)}$}
\par\end{center}

\textcolor{black}{Where }\textcolor{black}{\emph{$X$}}\textcolor{black}{{} is the system, $x$ is a state, and $P$ is a partition $P=M_{1}, \ldots ,M_{r}$.}

\textcolor{black}{Importantly, identifying the MIP requires searching all possible
partitions and comparing their effective information $\phi$. This effective information is specified in terms of effect and causal information, that is, the distance between two probability distributions: one for the unpartitioned (unconstrained) partition (this can be the full set of nodes of the whole system or one of its possible partitions) and a partition of this latter. Such probability distributions determine probabilities  of all possible future (effect) or past (causal) states of an arbitrary partition being in a current state. This means that comparing one set of nodes that can be the full set of nodes of the system or a subset (partition) of itself with all possible partitions of this set of nodes, MIP represents the partitions with the minimal value of the distance between probability distributions of the set of nodes and one of all its possible partitions.}

\textcolor{black}{When a set of nodes is chosen to compute effective information, this is referred to as a `mechanism', and the partition to which it is compared is referred to as the `purview'. The distance between probability distributions is computed by means of an adaptation of the Earth Mover's Distance (EMD) algorithm, which is a method to evaluate dissimilarity between two multi-dimensional distributions in a given feature space where a distance measure between single features, which we call the ground distance, is given. The EMD ``lifts'' this distance from individual features to full distributions. Note that EMD is referred to as a Wasserstein metric in mathematics, and is commonly used in machine learning as a natural metric between two distributions~\cite{villani}.}

\textcolor{black}{Intuitively, given two distributions, one can be seen as a mass of earth properly spread in space, the other as a collection
of holes in that same space. Then, the EMD measures the least amount of work needed to fill the holes with earth. Here, a unit of work corresponds to transporting (by an optimal transport method) a unit of earth a unit of ground distance.}
\section{Methods}
In this section we  introduction of the meta-perturbation analysis, additional technical details of which are presented in the Appendix. Next, we recap the causal perturbation and causal analysis leading up to the notion of program-size divergence, which is our core metric for how different programs, i.e. systems--more or less integrated--respond to perturbations
\subsection{\textcolor{black}{Programmability test and meta-test}}

\textcolor{black}{In \cite{zenilturingtest}a programmability test is introduced which was inspired by the Turing test, while being based on the view that the universe and all physical systems living in it and able to process information can be considered (natural) computers \cite{zenilturingtest} equipped with particular computational capabilities \cite{zenil2013behavioural}.}

\textcolor{black}{The programmability test is explained as: ``...replacing the question of whether a system is capable of digital computation with the question of whether a system can behave like a digital computer and whether a digital computer can exhibit the behaviour of a natural system.''}

\textcolor{black}{Then, in the same way that the Turing test proceeds
to ask questions of a computer in order to determine whether it is
capable of computing an intelligent behaviour, the programmability test aims to know what a specific system is capable of computing by means of algorithmic querying \cite{zenil2015causality}.}

\textcolor{black}{In practice, the programmability test is
a system perturbation test~\cite{zenilturingtest,zenilturingtest2} that "asks" questions of a computational system in the form: }\textcolor{black}{\emph{what
is your output (answer) given this question (input)?}}\textcolor{black}{.
This idea is applied to $\phi_{K}$'s implementation so that once the set of all possible answers of a system is obtained, this set is analyzed and generalized to deduce the rules that should not just offer a picture of its computability capabilities, but also simulate and give an account of the behaviour of the system itself.}

\textcolor{black}{A second step after this perturbation test is to analyze
its results in order to construct a computer program--as simple as possible--capable not only of reproducing the output repertoire but also of giving an account of the programmability capabilities of the system itself, that
is, rules capable of producing a certain output given an input, and at
the same time explaining where, in ordinal terms, such an output could be placed relative to the order of the full output repertoire. This latter aspect we refer to as the meta-perturbation test.}

\textcolor{black}{Then, $\phi_{K}$ not only applies a perturbation
test over a system, but also a meta-perturbation test over results
obtained on the first test. The rules found in this meta-test are
used not only as compressed specifications or representations of
the behaviour itself, but also as rules that give a sort of account
of the behaviour of the system. }

\textcolor{black}{This can be done because the systems analyzed in IIT
are well known, or in other words, since all node-by-node operations
are well defined, it is easy to compute all possible outputs (answers)
for all possible inputs (questions or queries), corresponding to what in IIT are referred to as repertoires. In the context of $\phi_{K}$, a meta-test is applied in order to find the rules that describe the behaviour embedded in repertoires of a system, instead of trying to ascertain the rules that define how the system works.}

\textcolor{black}{A system specified in this manner turns on a ``computer\textquotedbl{}, recording it's own behaviour (e.g the repertoires) as well as probing itself, e.g. the action of $\phi_{K}$, in such a manner as to potentially give an account of its own behaviour. To make
this possible a system specification must be enabled with an explanatory
interface based on these simple embedding behaviour rules. $\phi_{K}$
goes beyond the original $\phi$ in that the programmability test
searches for the rules underlying the behaviour of a system rather than
generating a description of its possible causal connections. While in
IIT these rules are defined a priori and induced by perturbation,
$\phi_{K}$’s objective is not only to find rules that simulate, but
also describe such behaviour in a brief manner (thus simple rules)
and make predictions about the behaviour of the system. The field
of Algorithmic Information Dynamics~\cite{maininfo,mainbook} implements
this approach by asking what changes to hypothesized outputs mean
for the hypothesized underlying program generating an observation,
after an induced or natural perturbation.}

\textcolor{black}{The simple rules discovered and used for the calculation
of $\phi_{K}$ are used here exclusively to compose }\textcolor{black}{\emph{constrained/unconstrained
distributions}}\textcolor{black}{{} used in IIT for obtaining }\textcolor{black}{\emph{cause-and-effect
information}}\textcolor{black}{, a key concept from which the integration
of information derives. The rest of the calculus--earth mover's distance
measurements, the calculus of conceptual spaces, major complex and
finding the MIP-- remains as specified in IIT 3.0.}

\subsection{Causal perturbation analysis}

From a statistical standpoint, it would be typical to suggest that the behaviour of two time series, let's call them $X$ and $Z$, would potentially be causally connected if they were statistically correlated. Yet, there are several other cases that would not be distinguishable after a correlation test. A first possibility is that the time series simply shows similar behaviour without being causally connected, i.e. there is a shared upstream causal driver $Y$, concealed from the observer. Another possibility is that they are causally connected, but that correlation does not tell us whether it is a case of $X$ affecting $Z$, or vice versa. 

\begin{figure*}[ht!]
\centering
\includegraphics[scale=0.3]{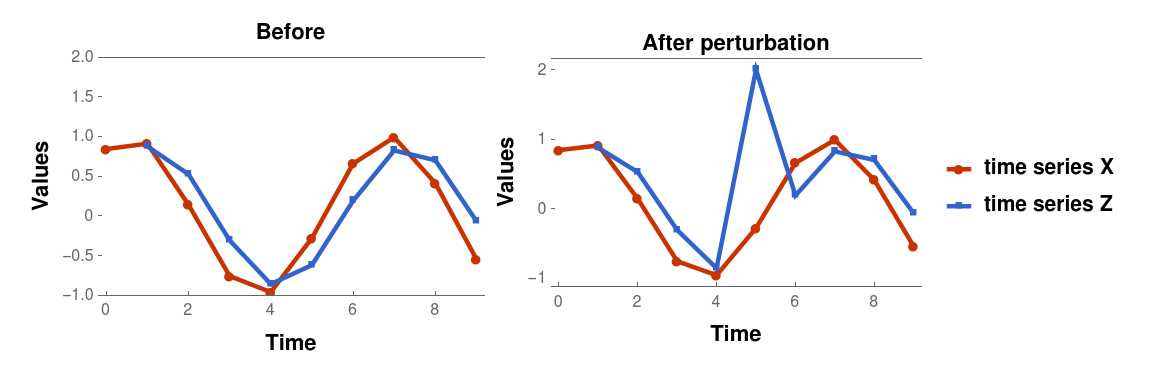}\\
\includegraphics[scale=0.3]{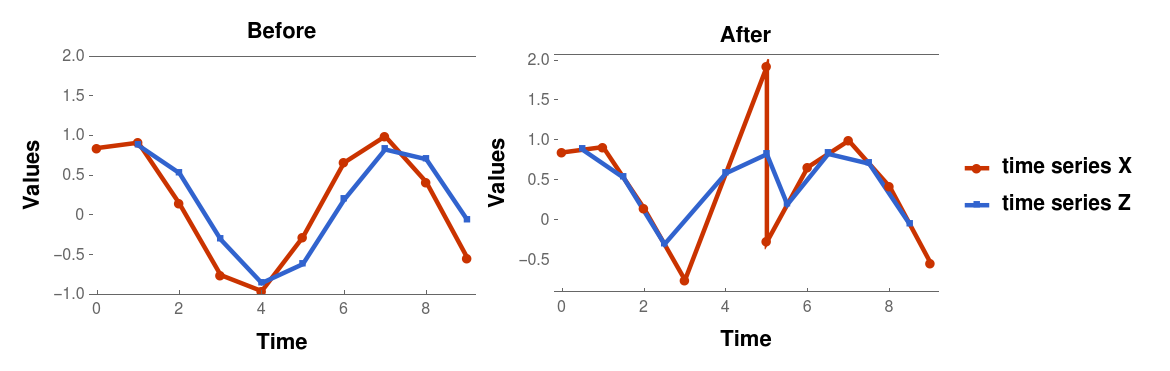}
\caption{\label{timeseriesbeforeafter}\textcolor{black}{Causal intervention analysis on time series $X$ and $Z$ before and after perturbation in $Z$ (top) and $X$ (bottom). The values of $Z$ come from the moving average of $X$, so there is a one-way causal relationship: perturbing $X$ has an effect on $Z$ but perturbing $Z$ has no effect on $X$ thereby suggesting the causal relationship.}}
\end{figure*}

Perturbation analysis allows some disambiguation. The idea is to apply a perturbation on one time series and see how the perturbation spreads to the other time series. Perturbing the data-point in position 5 the time series $Z$ as shown in Figure~\ref{timeseriesbeforeafter} multiplying it by -2, $X$ does not respond to the perturbation. This means that for this data point, $X$ remains the same. This suggests that there is no causal influence of $Z$ on $X$.

In contrast, if the perturbation is applied to a value of $X$, $Z$ changes and follows the direction of the new value, suggesting that the perturbation of $X$ has a causal influence on $Z$. From behind the scenes, we can reveal that $Z$ is the moving average of $X$, which means that each value of $Z$ takes two values of $X$ to calculate, and so is a function of $X$. The results of these perturbations produce evidence in favour of a causal relationship between these processes, if we did not know that they were related by the function we just described.

\begin{figure*}[htp]
\centering \includegraphics[scale=0.3]{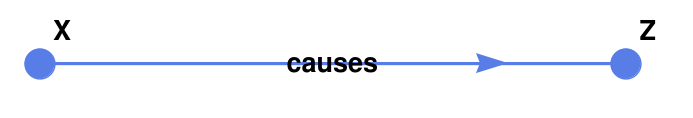} \caption{\label{causal1}Possible self-loopless causal relationship between two \textcolor{black}{unlabelled} variables
$X$ and $Z$.}
\end{figure*}

This suggests that it is $X$ which causally precedes $Z$. So we can say that this single perturbation suggests a causal relationship illustrated in Figure~\ref{causal1}.

\begin{figure*}[htp]
\centering \includegraphics[scale=0.25]{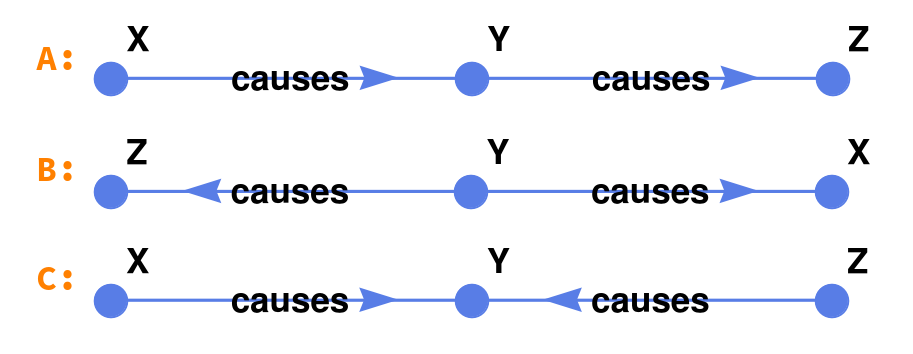} \caption{\label{timeseriesafterbefore2}Acyclic path graphs representing all possible \textcolor{black}{self-loopless}
connected causal relationships among 3 \textcolor{black}{ unlabelled} variables.}
\end{figure*}

There are a number of possible types of causal relationship between three events (see Figure~\ref{timeseriesafterbefore2}) that can be represented in what is known as a directed acyclic graph (DAG), that is, a graph that has arrows implying a cause and effect relationship but has no loops, because a loop would make a cause into the cause of itself, or an effect that is also its own cause, something that would be incommensurate with causality. In these graphs, nodes are events and events are linked to each other if there is a direct cause-and-effect relation.
In the first case, labelled $A$ in orange, the event $X$ is the cause of event $Y$, and $Y$ is the cause of event $Z$, but $X$ is said to be an indirect cause of $Z$. In general, we are, of course, always more interested in direct causes, because almost anything can be an indirect cause of anything else. In the second case $B$, an event $Y$ is a direct cause of both $Z$ and $X$. Finally, in case $C$, the event $Y$ has 2 causes, $X$ and $Z$. With an interventionist calculus such as the one performed on the time series above, one may rule out some but not all cases, but more importantly, the perturbation analysis offers the means to start constructing a model explaining the system and data rather than merely describing it in terms of simpler correlations.
In our approach to integrated information, \textcolor{black}{ the idea is to identify
the set of most likely generating candidates able to produce certain observed behaviour even if such behaviour may not carry any statistical regularity and for all purposes appear statistically random~\cite{devine2006ait}. Strictly speaking, computational mechanics~\cite{Shalizi2003Optimal,shalizi2001computational}, is a framework that bridges statistical inference and stochastic modelling that suggests a model based on an automaton called an $\epsilon$-machine. However, such machines are stochastic in nature and, if the methods used to reconstruct such machines rely on statistical methods, the result is only an apparent causal representation with no correspondence between internal states and alleged states of the phenomenon observed. In contrast, approaches based on algorithmic probability as approached by algorithmic information dynamics can complement computational mechanics as they provide means to construct non-stochastic automata models that are independent of probability distributions and are in a strict sense optimal and universal~\cite{solomonoff}.}

In the case of our two time series experiments, the time series $X$
is produced by the mathematical function $f(x)=Sin(x)$, and thus $Sin(x)$ is the generating mechanism of time series $X$. On the other hand, the generating mechanism of $Z$ is $MovAvg(f(x))$, and clearly $MovAvg(f(x))$ depends on $f(x)$, which is $Sin(x)$, but $Sin(X)$ does not depend on $MovAvg(f(x))$. In the context of networks, the algorithmic-information
dynamics of a network is the trajectory of a network moving in algorithmic-information
space together with the identification of those elements that shoot
the network towards or away from randomness.

\subsection{Causal influence and sublog program-size divergence}

According to Algorithmic Information Dynamics~\cite{maininfo,mainbook} there is an algorithmic causal relationship between two states $s_{t}$ and $s_{t'}$ of a system $M$ and $M'$ if

\[
|K(M_{s_{t}}) - K(M'_{s_{t'}})|\leq \log_{2}(t)+c
\]

That is, if the descriptions of such systems can be bounded by $\log_{2}$ and a small constant $c$, then $M$ is most likely equal to $M'$ but in some other time state.

In other words, if there is a causal influence of $s_{t}$ on $s_{t+1}$ or $s_{t+1}$ on $s_{t}$ as a system in isolation, their $M$ and $M'$ short descriptions should not differ by more than the description of the difference.

\textcolor{black}{However, if the descriptions of the states of a system (which may be two systems) in different alleged state times are not causally connected, their difference will diverge beyond above bound. In integrated information, causal influence among its parts is what is claimed to be measured and how different elements of a system can be explained by a single model or the other parts of the system informs us as to how integrated a system may be. A system characterized by large divergence is less integrated compared to a system which evolves with small differences in its respective subpart descriptions.}

We will suggest that perturbations have to be algorithmic in nature because they need to be made or quantified at the level of the generating mechanisms from the whole or different parts of the integrated system and not at the level of the observations. For example, some $n-$ary expansions of the mathematical constant $\pi$ according to BPP (named after Bailey-Borwein-Plouffe) formulas~\cite{bpp} allow perturbations to the digits that do not have any further effect because no previous digits are needed to calculate any other segment of $\pi$ in the same base. The constant $\pi$ then can be said to be information disintegrated to the extent of the BPP representations. Algorithmically low complexity objects have low integrated information. Similarly, highly random systems have low integrated information, because perturbations have little to no impact. Integrated information is, therefore, a measure of sophistication that separates high integration from both random and trivially non-random states.

\subsection{A simplicity versus complexity test}

With the previous section in mind we can proceed to introduce the idea of $\phi_{K}$, where $K$ stands for the letter often used for algorithmic (from Kolmogorov or Kolmogorov-Chaitin) complexity, and $\phi$ for the traditional of integrated information theory~\cite{oizumi2014phenomenology}. The measure $\phi_{K}$ mostly follows methods that Oizumi and Tononi set forth in \cite{oizumi2014phenomenology}, where integrated information is
measured, roughly speaking, as distances between probability distributions that characterize a MIP (Minimum Information Partition), that is, ``the partition of [a system] that makes the least difference''\cite{oizumi2014phenomenology}.

\textcolor{black}{However, the difference between IIT's $\phi$ and
$\phi_{K}$ lies in how $\phi_{K}$ circumvents what is called the
``intrinsic information bottleneck principle’’~\cite{oizumi2014fromthe}
that traditionally requires an exhaustive search for the MIP among
all possible partitions of a system, a procedure responsible for the
fact that integrated information computation requires super-exponential
computational resources. In contrast to $\phi$, which follows a statistical
approach to estimating and exhaustively reviewing repertoires, the approach
to $\phi_{K}$ is based on principles of algorithmic information.}

\textcolor{black}{Discovering the simple rules that govern a ``discrete
dynamical system'' \cite{mayner2017pyphia} like those studied in
IIT presupposes the analysis
of its general behaviour in pursuit of a dual agenda: first, to determine its computational capabilities, and secondly to obtain explanations and descriptions of the behaviour of the system.}

\textcolor{black}{As a consequence, one of the major adaptations of IIT is that $\phi_{K}$ uses the concept of Unconstrained Bit Probability Distribution (UBPD), that is, the individual probabilities associated with a node of a system taking values of 1 (ON) or 0 (OFF) after it has been ``fed'' all its possible inputs or after a perturbation.}

\textcolor{black}{In the context of $\phi_{K}$, UBDP is computed using
simulation and definition systems governed by simple rules, unlike $\phi$, which uses the TPM (Transition Probability Matrix) to compute
IIT's unconstrained/constrained probability distributions.}

\textcolor{black}{In Figure 4 and Table 1 the concept UBPD and its calculus
is explained, using the example used by Oizumi et. al. in~\cite{oizumi2014phenomenology}.}

\begin{table}[ht]
\begin{lstlisting}
am = {{0, 1, 1}, {1, 0, 1}, {1, 1, 0}}; 
dyn = {"OR", "AND", "XOR"}; 
calcUBPOutputs[1,am,dyn]//AbsoluteTiming 
calcUBPOutputs[2,am,dyn]//AbsoluteTiming 
calcUBPOutputs[3,am,dyn]//AbsoluteTiming
\end{lstlisting}

\begin{lstlisting}
{0.000575,<|"ZeroProb"->0.25,"OneProb"->0.75|>}
{0.000287,<|"ZeroProb"->0.75,"OneProb"->0.25|>}
{0.000341,<|"ZeroProb"->0.5,"OneProb"->0.5|>}
\end{lstlisting}

\caption{Computing UBPD for system shown in Figure 1. Lines 1, 2: Definition of the system in Figure 1 by adjacency matrix (line 1) and dynamics (line 2). Line 3-5: Calculation of individual probabilities that each node of the system will take values 0/1 across the whole output repertoire. Results square: time of computation in seconds and UBPD distribution. }
\end{table}

\textcolor{black}{In order to explain the notion of UBPD, in Figure 4 we use Oizumi's example used in~\cite{oizumi2014phenomenology} to calculate information integration. Figure 4-A shows the network representation: three nodes fully connected with different types of operation executed on its inputs, that is, for example, inputs to node A (coming from B and C nodes) will be processed in a logical OR operation. In Figure 4-B the adjacency matrix that represents the same same network is shown. This adjacency matrix uses the number 1 to indicate if a node receives signals (inputs) for another node. For example, the first row in the adjacency matrix indicates that node 1 or A receives inputs from nodes B and C, denoted as nodes 2 and 3. Finally, Figure 4-C shows the full input and output repertoires, that is, for the full set of all possible inputs to this system, all corresponding outputs are calculated according to the logical operations
defined.}

\textcolor{black}{Table 1 shows code for computing UBPD for the system in Figure 4. This computation starts with the specification of
the adjacency matrix (line 1) and internal dynamic (line 2) of the
target system. Then, lines 1 and 2 in Table 1 represent code to network specified in Figures 4-A and 4-B.}

\textcolor{black}{In the IIT approach, the system is perturbed with all possible inputs to obtain the full output repertoire (Figure 4-C).
Then, in the context of $\phi_{K}$, UBPD corresponds to the distribution of probabilities that each node will take values 0/1 in the output/input
repertoires after the perturbation. For instance, in Figure 5-A,
full input and output repertoires are shown for network in Figure 4-A.
Now, let's say we want to compute the future probability distribution,
that is, the probability necessary to compute effect information according
to \cite{oizumi2014phenomenology}. In this case we take output repertoire as a reference and we compute the probability of nodes in the future (outputs) taking the values 0 or 1. For node A, for example, the probability that node A takes the value of 1 is 0.25, that is 2/8, and that it takes the value
of 0 is 0.75 or 6/8. These values are called the UBPD for node A. A resume of UBPD for all nodes is given in Figure 5-B.}

\textcolor{black}{Once UBPD is computed for a subject partition, in this case the full system's probability distribution is computed by multiplying UBDPs. Let's take as an example the future probability of input \{0,
0, 0\}, computed as $P(A) = 0$ {*} $P(B) = 0$ {*} $P(C) = 0$, that is, 0.25 {*} 0.75 {*} 0.5 (see first row in Figure 5-C). When all future probabilities are computed in this manner, the result is the distribution shown in Figure 5-D, which is exactly the same one computed in \cite{oizumi2014phenomenology},
as shown in Figure 5-E.}

\textcolor{black}{In general, UBDP is used to compute probability distributions
of a system in the context of $\phi_{K}$, which mirrors the ``constrained/unconstrained
probability distributions'' in \cite{oizumi2014phenomenology}, that is, probability distributions of input/output patterns for specific configurations (partitions) of the system, in contrast to what IIT 3.0 does. In this last case, Mayner shows how probability distributions are computed in the context of IIT in his S1 text mentioned in \cite{mayner2017pyphia},
using terms such as ``marginalization'' and ``virtual
elements'' that seem to be highly complex methods.}

\textcolor{black}{Then, in the context of $\phi_{K}$, UBDP aims to obtain the same results in terms of probability distributions, in a manner equivalent to IIT but by following a different conceptual
approach. Our measure $\phi_{K}$ uses adapted methods, having
algorithmic complexity as a background, to compute information integration.}

\textcolor{black}{In Table 1, lines 3-5 shows Mathematica code that
computes UBDP for the system specified in lines 1 and 2, that is, by means of an adjacency matrix and an array of computations that nodes perform, or the system dynamics. Table 1 also shows results of this computation in this order: 1) time needed to compute, followed by probability that a node take the value zero (zeroProb) or the value one(oneProb).} One can see how the results in Table 1 correspond to UBDP values shown in Figure 5-B.

\begin{figure}
\begin{centering}
\includegraphics[scale=0.44]{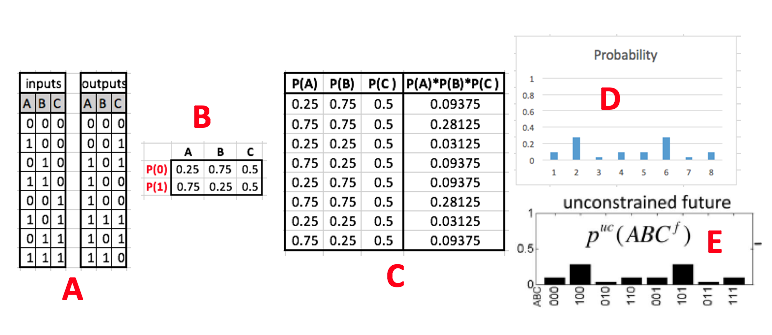}
\par\end{centering}
\caption{An example of using UBPC to calculate unconstrained output distribution.}
\end{figure}
\textcolor{black}{We should note that for $\phi_{K}$, computation
of probability distributions seems to be a task of counting, which for huge systems would be extremely difficult or even impossible, if attempted in a classical/brute force way. But, two important facts should be pointed out here: 1) In the context of $\phi_{K}$, UBPD is not calculated in this traditional way, but is calculated using the simulation of the behaviour of a system represented by a set of simple rules. Then for $\phi_{K}$, an exhaustive review of repertoires is not needed to compute the individual probabilities shown in Table 1, and 2) despite strong theoretical and methodological differences between them, $\phi_{K}$ and $\phi$ lead to the same
results.}

In the next sections we derive simple rules of a system, using the perturbation test and its application to implement $\phi_{K}$.

\section{Numerical Results}

\subsection{Compression sensitivity as informative of integration}

To understand the relationship between IIT and algorithmic complexity, we shall briefly move away from the case of networks and focus on binary files and the binary programs that may generate them, the programs that are natural computable candidate models explaining the data. To illustrate the connection, let us take some extreme cases. Let's say we have a random file:

\begin{lstlisting}[language=Mathematica, caption=]

In: = randomfile=RandomChoice[CharacterRange["a", "z"],100]
Out = {l,q,l,d,d,x,f,e,u,l,u,b,m,y,a,l,b,l,v,a,v,v,t,l,h,o,x,d,i,a,z,t,z,b,r,y,v,h,c,n,s,u,l,u,w,w,e,a,i,h,h,w,f,d,d,k,o,c,a,k,u,x,v,n,v,e,c,r,t,c,g,s,g,x,y,t,c,h,k,w,c,t,y,u,e,k,y,v,a,h,t,k,f,y,c,r,b,y,y,x}

In: = Compress[randomfile]
Out = 1:eJx1kEEOgyAQRWnSi/Q+XbUnEAUhtJoKInp6zbB5CenmZTL8mf+
Hh55fdlBKxfuFp4/pfbuKj/CHegCL0AqNcIWy1lr4Fe7CDhqNOuM1g0zihDP
caxKP2QNTB1wWZMjY1gsnYfxzxQYaeHnscdDY5q8CkvfYEOBSkG1CbTC14L
raGZF8xJ69UTo4bugn6Fc4hubHOuxJ0FgomVOjX1lOUdJPvg==

In: = Length[Characters@Compress[randomfile]]
Out = 222
\end{lstlisting}

\textcolor{black}{So, using the Compress
algorithm, the resulting compressed object is even longer, this is because the compression algorithms inserts the decompression instructions together with the checksum which ends up increasing the size of the resulting object if the object was not long and compressible enough to begin with.}

This is what happens if we perform a couple of random perturbations to
the uncompressed file:

\begin{lstlisting}[language=Mathematica, caption=]
In: = mutatedfile=ReplacePart[randomfile,{5->"k", 12->"x"}]
Out = {l,q,l,d,k,x,f,e,u,l,u,x,m,y,a,l,b,l,v,a,v,v,t,l,h,o,x,d,i,a,z,t,z,b,r,y,v,h,c,n,s,u,l,u,w,w,e,a,i,h,h,w,f,d,d,k,o,c,a,k,u,x,v,n,v,e,c,r,t,c,g,s,g,x,y,t,c,h,k,w,c,t,y,u,e,k,y,v,a,h,t,k,f,y,c,r,b,y,y,x}
\end{lstlisting}

The difference between the original and perturbed files is:

\begin{lstlisting}[language=Mathematica, caption=]

In := SequenceAlignment[randomfile,mutatedfile]//Column
Out= {l,q,l,d} {{d},{k}} {x,f,e,u,l,u} {{b},{x}} {m,y,a,l,b,l,v,a,v,v,t,l,h,o,x,d,i,a,z,t,z,b,r,y,v,h,c,n,s,u,l,u,w,w,e,a,i,h,h,w,f,d,d,k,o,c,a,k,u,x,v,n,v,e,c,r,t,c,g,s,g,x,y,t,c,h,k,w,c,t,y,u,e,k,y,v,a,h,t,k,f,y,c,r,b,y,y,x}
\end{lstlisting}

The files only differ by 2 characters, which can be counted using the following code:

\begin{lstlisting}[language=Mathematica, caption=]
In := Total[Length/@First/@Select[SequenceAlignment[randomfile,mutatedfile],Head[\#[[1]]]==List\&]]
\end{lstlisting}

That is, 2/100 or 0.02 percent.\\
 
On the other hand, let's take a simple object consisting of the repetition of a single object, say the letter e:

\begin{lstlisting}[language=Mathematica, caption=]
In := simplefile=Table["e",100] 
Out = {e,e,e,e,e,e,e,e,e,e,e,e,e,e,e,e,e,e,e,e,e,e,e,e,e,e,e,e,e,e,e,e,e,e,e,e,e,e,e,e,e,e,e,e,e,e,e,e,e,e,e,e,e,e,e,e,e,e,e,e,e,e,e,e,e,e,e,e,e,e,e,e,e,e,e,e,e,e,e,e,e,e,e,e,e,e,e,e,e,e,e,e,e,e,e,e,e,e,e,e}
\end{lstlisting}

A shortest program to generate such a file is just: 

\begin{lstlisting}[language=Mathematica, caption=]
Table["e", 100]
\end{lstlisting}

In other languages this could be produced by an equivalent `For' or `Do-While' program. We can now perturb the program again, without loss of generality. Let's allow the same 2 perturbations to the data only, and not to the program instructions (we will cover this case later). The only places that can be modified are thus `e' or 1 instead of 5, say: Table[``a'',500] 

\begin{lstlisting}[language=Mathematica, caption=]
In := Length/@First/@Select[SequenceAlignment[Table["a",500],Table["e",100]],Head[#[[1]]]==List&] 
Out = {500} 
\end{lstlisting}

Now, the original and decompressed versions differ by 500 elements, and not just a small fraction (compared to the total program length) as in the random case. This will happen in the general case with random and simple files; random perturbations will have a very different effect on each case.

An object that is highly integrated among its parts means that one can explain or describe part of each part with some other part when the object is algorithmically simple; then these parts can be compressed by exploiting the information that the said other parts carry over from yet others, and the resulting program will be highly integrated only if the removal of any of these parts has a non-linear effect on its generating program. In a random system, no part contains any information about any other, and the distribution of the individual algorithmic-content contribution of each element is a normal distribution around the mean of the algorithmic-content contributions, hence poorly integrated and trivial. So integrated information is a measure of sophistication, filtering out simple and random systems, and only ascribing high algorithmic information content to highly integrated information systems.

The algorithmic information calculus thus consists of a 2-step procedure to determine:

\begin{enumerate}
    \item The complexity of the object (e.g. string, file, image) \item  The elements in that object that are less, more, or not sensitive to perturbations that can `causally steer the system,' i.e. causally modify an object in a surgical algorithmic fashion rather than on the basis of guesswork based on statistics.
\end{enumerate}

Note that this causal calculus is semi-computable, and one can perform guiding perturbations based upon approximations ~\cite{maininfo,mainbook}. Also note that we did not cover the case in which the actual instructions of the program were perturbed. This is actually just a subcase of the previous case, that separates data from program. For any program and data, however, we conceive an equivalent Turing machine with empty input, thus effectively embedding the data as part of its instructions.
Nevertheless, the chances of modifying the instruction Print[] in the random file case are constant, and for the specific example are: $7/107 = 0.0654$. While for the non-random case, the probability
of modifying any piece of the Table[] function is: $8/12 = 0.666667$. Thus, the break-up of a program of a highly causally generated system is more likely under random perturbations. 

Notice similarities to a checksum for, e.g., file exchange verification (e.g. from corruption or virus infection for downloading from the Internet), where the data to be transmitted is a program and the data block to check is the program's output file (which acts as a hash function).

Unlike regular checksums, the data block to check is longer than the program, and the checking is not for cryptographic purposes. Moreover, the dissimilarity distance between the original block (shared information)
and the output of the actual shared program provides a measure of both how much the program was perturbed and the random or nonrandom nature of the data compressed by the program. And just like checksums, one cannot tamper with the program without perturbing the block to be verified (its output), without significantly changing the output (block) if what the program has encoded is nonrandom and therefore causally/recursively/algorithmically generated. Of course all the theory is defined in terms of binary objects, but for purposes of illustration and with no loss of generality we have shown actual programs operating on larger alphabets (ASCII). And we also decided to perform perturbations on what seems to be the program data and not the program itself (though we have seen that this distinction is not essential) for illustration purposes, to avoid the worst case in which the actual computer program becomes non-functional.

Yet, this means that the algorithmic calculus is actually more relevant, because it can tell us which elements in the program break it completely and which ones do not. But what happens when changes are made to the program output and not the program instructions? Say we exchange an arbitrary e for an a in our simple sequence consisting of a single letter, e.g. the third entry ('a' for `e'):

If we were to look to the generating program of the perturbed sequence,
this would need to account for the a, e.g. 

\begin{lstlisting}[language=Mathematica, caption=]
In := ReplacePart[Table["e",100],3->"a"]
Out = {e,e,a,e,e,e,e,e,e,e,e,e,e,e,e,e,e,e,e,e,e,e,e,e,e,e,e,e,e,e,e,e,e,e,e,e,e,e,e,e,e,e,e,e,e,e,e,e,e,e,e,e,e,e,e,e,e,e,e,e,e,e,e,e,e,e,e,e,e,e,e,e,e,e,e,e,e,e,e,e,e,e,e,e,e,e,e,e,e,e,e,e,e,e,e,e,e,e,e,e}
\end{lstlisting}

\noindent where the second program is longer than the original one, and has to be, if the sequence is simple, but the program remains unchanged if the file is random because the shortest program of a random sequence
is the random sequence itself, and random perturbations keep the sequence random. Furthermore, every element in the simple example consisting of repetitions of e has exactly the same algorithmic content contribution when changed or removed, as all programs after perturbation are of the form: 

\begin{lstlisting}[language=Mathematica, caption=]
ReplacePart[Table["e",100],n->x]
\end{lstlisting}

Notice also how this is related to $\phi$ and possibly any measure of integrated information based on the same principles. 

We can now apply all these ideas to the language of networks, with respect to which IIT has, for the most part, been defined. We have shown before that networks with different topologies have different algorithmic complexity values~\cite{physicaa}, in accordance with the theoretical expectation. In this way, random ER graphs, for example, display the highest values, while highly regular and recursive graphs display the lowest~\cite{zenilreview}. Some more probabilistic, but yet recursively generated graphs are located between
these 2 extremes~\cite{zenilmethods}. Indeed, the algorithmic complexity $K$ of a regular graph grows by $O(logN)$, where $N$ is the number of nodes of the graph, as in a highly compressible complete graph. Conversely, in a truly random ER graph, however, $K$ will grow by $O(\log E)$, where $E$ is the number of edges, because the information about the location of every edge has to be specified.

In what follows we will perform some numerical tests strengthening our analytic derivations.

\subsection{Finding simple rules in complex behaviour}

A perturbation test is applied to systems which IIT is interested
in. The set of answers is analyzed in order to find the rules that
1) make it possible to simulate the behaviour of the system, 2) define their computability power, that is, rules that give an account of what the system can and cannot compute, and 3) rules able to describe and predict behaviour of the same system. The following procedure was applied to estimate $\phi_{K}$. 

\begin{enumerate}
\item The perturbation test was applied to systems used in IIT to obtain detailed behaviour of the systems. 
\item Results in step one were analyzed in order to reduce the dynamics of a system to a set of simple rules. That is, in keeping with the claims of  natural computation, we found simple rules to describe a system's behaviour. 
\item Rules found in step 2 were used to generate descriptions of what a system
is or is not capable of computing and under what initial conditions,
without having to calculate the whole output repertoire. 
\item A combination of rules found in steps 2 and 3 was used to develop
procedures for predicting the behaviour of a system, that is, whether it is possible to have reduced forms that express complex behaviour. Knowing what conditions are necessary for the system to compute something, it is possible to pinpoint where in the whole map of all possible inputs (questions) of a system such conditions may be found. 
\item Once rules in steps 2 and 3 are formalized, $\phi_{K}$ was turned into a kind of interrogator whose purpose was to ask questions of a system about its own computational capabilities and behaviour.
\end{enumerate}

This kind of analysis allowed us to find that the information distribution in the complex behaviour of systems analyzed in IIT followed a \textcolor{black}{distribution
replicated at several scales that is usually and informally identified as `nested' or `fractal', and means that it is susceptible of being summarized in simple rules by iteration or recursion, just as is the case with fractals proper. These properties are used to find compressed forms to
express answers given by a system when asked for explanations of its
own behaviour.}

This means that, as noted before, $\phi_{K}$ does not compute the
whole output repertoire for a system but uses simple rules to express
the whole behaviour of the system. Interestingly, the way in which
we proceed appears to be connected to whether or not the system itself
can explain its behaviour, or rather whether it can see itself to be
capable of producing its behaviour from an internal experience (configuration)
which is then evaluated by an observer. So $\phi_{K}$ takes the form
of an automatic interrogator that, in imitation of the perturbation
test, asks questions of the form \emph{are you capable of this specific
configuration? (pattern), and if so, say where, in the map of the
behavioural repertoire, I can find it}.

The benefit of representing systems using simple rules is that it allows an alternative calculation closer to algorithmic complexity and the potential to reduce the number of calculations to derive an educated estimation as compared to the original version of IIT 3.0.

At this point, it is not possible to explain how simple rules define
a system in the context of $\phi_{K}$ without talking about the pattern
of distribution of information in the behaviour of systems like those studied in IIT. 

\subsection{Simple rules and the \textcolor{black}{pattern of }distribution of
information}

As shown in~\cite{zenil2015causality}, despite deriving from a very
simple program, without knowing the source code of the program, a
partial view and a limited number of observations can be misleading
or uninformative, illustrating how difficult it can be to reverse-engineer
a system from its evolution in order to discover its generating code~\cite{zenil2015causality}.

In the context of IIT, when we talk about a complex network we find
that there are different levels of understanding complex phenomena,
such as knowing the rules implemented by each node in a system and finding
the rules that describe its behaviour over time. To achieve the second,
as perhaps could be done for the ``whole {[}of{]} scientific practice''
\cite{zenil2015causality}, we found it useful to perform perturbation tests in order to deduce the behaviour of the subject systems. Results
were analyzed and a \textcolor{black}{pattern in the }distribution of
information was found to characterize the behaviour of these kinds of systems. Then, as was to be expected, replicating behaviours were
amenable to being expressed with simple formulae.

In order to explain how simple rules were found and implemented in
$\phi_{K}$, consider as an example the 7-node system shown in Figure
6 whose behaviour is computed by perturbing the system on all possible inputs. The results, or the whole output repertoire, is shown in Table 6 in
Appendix A.

\begin{figure}[ht]
\begin{centering}
\includegraphics[scale=0.5]{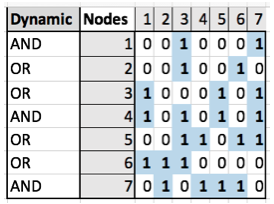} \includegraphics[scale=0.5]{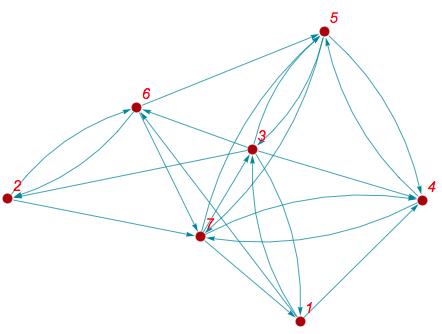} 
\par\end{centering}
\caption{Seven-node system. Adjacency matrix and network representation.}
\end{figure}

The strategy adopted to find rules that govern a system's behaviour is the same used in almost any branch of science, which is to say we separately observe the behaviour of some of the components of a phenomenon, in this case nodes, while bearing in mind that this behaviour is not isolated but rather the by-product of interacting elements, or in other words, we observe individual behaviours without losing sight of the whole.

\textcolor{black}{When we observe the whole behaviour of the system shown in Figure 6 (see Appendix A, Table 6), we notice mostly chaotic behaviour but with subtle repetitions of certain patterns}. 

\textcolor{black}{When the behaviour of elements is isolated, the picture appears clearer. For example, Table 2 shows the isolated behaviour of nodes \{4\} and \{5\} of the same subject system}.

\begin{table}
\begin{lstlisting}
analysis07[[All, 4]]
\end{lstlisting}

\begin{lstlisting}
{0, 0, 0, 0, 0, 0, 0, 0, 0, 0, 0, 0, 0, 0, 0, 0, 0, 0, 0, 0, 0, 0, 0, 0, 0, 0, 0, 0, 0, 0, 0, 0, 0, 0, 0, 0, 0, 0, 0, 0, 0, 0, 0, 0, 0, 0, 0, 0, 0, 0, 0, 0, 0, 0, 0, 0, 0, 0, 0, 0, 0, 0, 0, 0, 0, 0, 0, 0, 0, 0, 0, 0, 0, 0, 0, 0, 0, 0, 0, 0, 0, 0, 0, 0, 0, 1, 0, 1, 0, 0, 0, 0, 0, 1, 0, 1, 0, 0, 0, 0, 0, 0, 0, 0, 0, 0, 0, 0, 0, 0, 0, 0, 0, 0, 0, 0, 0, 1, 0, 1, 0, 0, 0, 0, 0, 1, 0, 1}
\end{lstlisting}

\begin{lstlisting}
analysis07[[All, 5]]
\end{lstlisting}

\begin{lstlisting}
{0, 0, 0, 0, 1, 1, 1, 1, 1, 1, 1, 1, 1, 1, 1, 1, 0, 0, 0, 0, 1, 1, 1, 1, 1, 1, 1, 1, 1, 1, 1, 1, 1, 1, 1, 1, 1, 1, 1, 1, 1, 1, 1, 1, 1, 1, 1, 1, 1, 1, 1, 1, 1, 1, 1, 1, 1, 1, 1, 1, 1, 1, 1, 1, 1, 1, 1, 1, 1, 1, 1, 1, 1, 1, 1, 1, 1, 1, 1, 1, 1, 1, 1, 1, 1, 1, 1, 1, 1, 1, 1, 1, 1, 1, 1, 1, 1, 1, 1, 1, 1, 1, 1, 1, 1, 1, 1, 1, 1, 1, 1, 1, 1, 1, 1, 1, 1, 1, 1, 1, 1, 1, 1, 1, 1, 1, 1, 1}
\end{lstlisting}

\caption{Isolated outputs for nodes \{4\} and \{5\} in system introduced
in Figure 6 after perturbation.}
\end{table}

\textcolor{black}{In Table 2, the isolated behaviour of two nodes of the system in Figure 6 is shown, where it is possible to observe that isolated behaviours for \{4\} and \{5\} follow a sort of order. Such patterns are summarized in what we call behaviour tables, shown in Figure 7.}

\begin{figure}
\begin{centering}
\includegraphics[scale=0.27]{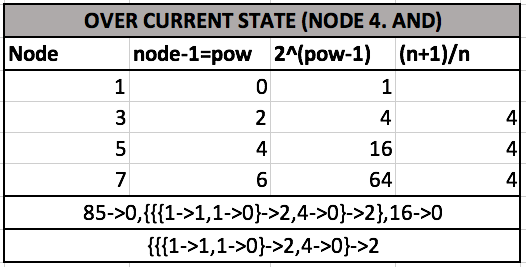}
\includegraphics[scale=0.3]{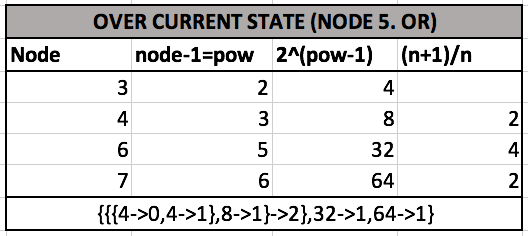} 
\par\end{centering}
\centering{}\caption{Behaviour Tables for seven-node system in Figure 6. (A) node 4, (B) node 5. From left to right, \textbf{Node} column lists input-nodes that feed the target node. \textbf{$node-1=power$} column computes the power used to transform a pattern in the world of the 7-node systems from binary to decimal. \textbf{2\textasciicircum pow} column is the result of the binary to decimal transformation operation. The fourth column contains divisions between elements indexed by $n+1$ in column 3 divided by the element indexed as $n$}
\end{figure}

Lowest rows in behaviour tables shown in Figure 7 (within braces) correspond to a compressed representation of behaviours shown in Table 2.

Compressed expressions of behaviour for node \{4\}, for instance, means: 85 repetitions of digit 0, followed by a pattern repeated
2 times, this pattern being: 1 once, followed by one 0 (that is \{1->1, 1->0\}). This last pattern is followed for four digits 0, and so on. Here notice that the first number 85 is the sum of the numbers shown in the 3rd column of its behaviour table.

For node 5, a compressed representation of the behaviour means: four times digit 0, followed by four digits 1. This pattern is followed by eight repetitions of digit 1. The last pattern is followed
for 94 ($32+64$) repetitions of digit 1.

The representation used in this isolation of behaviours is expressed in terms of the nodes that ``feed" into target nodes of this example (Node column), namely nodes \{4, 5\} whose inputs, according to Figures 6 and 7 are: for node \{4\}: \{1,3,5,7\}, and for node 5: \{3,4,6,7\}.

\textcolor{black}{This first shallow analysis works to yield the intuition that the behaviour of an isolated node can be expressed as a series of regularities
in terms of its inputs. In this context, intuition tells us that
the greater the number of regularities, the shorter the description;
then if no patterns are detected the chances of a causal relationship
are lower.}

\textcolor{black}{Perspective changes when rule/algorithm or compressed expression of behaviour is not constructed from regularities identified at first sight, but from intrinsic algorithmic properties. In this latter case, behaviour of systems can be expressed as patterns of information with a distribution
replicable at different scales, what we here call  \emph{fractal representation}
or \emph{fractal behaviour}. To explain what we mean by fractal, we introduce characteristics of distribution of information for the 7-node system shown in Figure 6 analyzed using $\phi_{K}$. This implementation is shown in Table 3.}

\begin{table}
\begin{lstlisting}
cm07 = {{0, 0, 1, 0, 0, 0, 1},          
{0, 0, 1, 0, 0, 1, 0},
{1, 0, 0, 0, 1, 0, 1},
{1, 0, 1, 0, 1, 0, 1},
{0, 0, 1, 1, 0, 1, 1},
{1, 1, 1, 0, 0, 0, 0},
{0, 1, 0, 1, 1, 1, 0}};
dyn07 = {"AND", "OR", "OR", "AND", "OR", "OR", "AND"};
(*computing places in output repertoire where node 4 = 0*)
res070=onPossibleBehaviour[{4}, {0}, dyn07, cm07] 
(*Summarized representation of fractal behaviour*)
gp=givePlaces[res070["DecimalRepertoire"],res070["Sumandos"]];
\end{lstlisting}

\begin{lstlisting}
<|"DecimalRepertoire"-> {0, 1, 4, 5, 16, 17, 20, 21, 64, 65, 68, 69,80, 81, 84}, 
"Sumandos"-> {0, 2, 8, 10, 32, 34, 40, 42}|>
\end{lstlisting}

\begin{lstlisting}
{0, 1, 2, 3, 4, 5, 6, 7, 8, 9, 10, 11, 12, 13, 14, 15, 16, 17, 18, 19, 20, 21, 22, 23, 24, 25, 26, 27, 28, 29, 30, 31, 32, 33, 34, 35, 36, 37, 38, 39, 40, 41, 42, 43, 44, 45, 46, 47, 48, 49, 50, 51, 52, 53, 54, 55, 56, 57, 58, 59, 60, 61, 62, 63, 64, 65, 66, 67, 68, 69, 70, 71, 72, 73, 74, 75, 76, 77, 78, 79, 80, 81, 82, 83, 84, 86, 88, 89, 90, 91, 92, 94, 96, 97, 98, 99, 100, 101, 102, 103, 104, 105, 106, 107, 108, 109, 110, 111, 112, 113, 114, 115, 116, 118, 120, 121, 122, 123, 124, 126}
\end{lstlisting}

\caption{$\phi_{K}$ asking for accounts of information distribution in
behaviour of 4th node of the system shown in Figure 6. \textbf{Lines
1-8}: Definition of the 7-node system by means of adjacency matrix and
its internal dynamics. \textbf{Line 10:} $\phi_{K}$'s code asking
for zero digit location in the whole behaviour of node 4. \textbf{Line
12:} Compressing answer given by the system in line 10. \textbf{Lines
1 and 2 in results square:} Compressed form of the 0 digit distribution
in the behaviour of node 4. The second grey square above shows the unfolded
answer of the system.}
\end{table}

\textcolor{black}{Table 3 shows how behaviour of the system shown in Figure 6 can be expressed
as simple rules following an analysis based on a querying scheme that results in a reduced form to express its information distribution
as a pattern replicated at different scales or as a fractal form.
Answers given by systems join facts explored above on regularities and the fractal distribution of information. It is important to note 
that the querying scheme has to be computable and algorithmically
random in order to avoid introducing an artificially random-looking
behaviour from the observer (experimenter/interrogator) to the observed
(the system in question).}

\textcolor{black}{In Table 3, after defining the target system by means of an adjacency
matrix and a dynamics vector (lines 1 to 8), $\phi_{K}$ can be regarded
as testing: \emph{how 0 is distributed in node 4 in the system of
seven nodes }(line 10).}

\textcolor{black}{The target system reacts to the $\phi_{K}$'s query and it ``answers''
in a compressed form (Table 3, second part, lines 1 and 2). The result
can be represented in compressed form, expressed as a tiny rule that
represents what we have called a fractal pattern. Such an expression is defined,
as can be seen in the second square in Table 3, by two variables: \emph{DecimalRepertoire} that holds points distanced in different proportions where the patterns defined by the \emph{Sumandos} variable must be reproduced. This means that in order to unfold the whole distribution (of digit zero), the
pattern of numbers in \emph{Sumandos} must be added to each value
in \emph{DecimalRepertoire}.}

\textcolor{black}{Once this `fractal' simple rule is unfolded, we obtain the ordinal
places where, in the whole behaviour of node 4, digit 0 can be
found (see third square in Table 3). The accuracy of this answer can
be verified by counting ordinals where, for node 4, its output = 0 in
Table 6 in Appendix A, taking into account that counting starts at 0.}

In summary, $\phi_{K}$ is turned into a kind of interrogator that asks a system about its own behaviour. On the other hand, a system is implemented as a set of rules that answers in different ways, depending on the information requested. This is unlikely with traditional approaches to $\phi$, whose representation of the system consists of the whole output repertoire of the system, which might represent an important disadvantage when large networks are analyzed. $\phi_{K}$'s answers use compressed forms taking
advantage of the fractal distribution of the information in the behaviour of the system, for which the answering interface is a function of its input related to each node in question.

Obviously the whole behaviour of a system is not about isolated elements, but about elements interacting in a non-linear manner, as IIT 3.0 makes clear. This last, broader view is also addressed in  terms of $\phi_{K}$, and explained in the following sections. In the next one, the advantages of simple rules over classical/naive approaches based on an exhaustive calculus and review of whole repertoires held in memory will be established.

\subsubsection{Automatic meta-perturbation test}

It can be seen that this querying system is similar to the programmability tests suggested in~\cite{zenilturingtest,zenilturingtest2,maininfo} based on questions designed to ascertain the programmability of a dynamical system.

The last section shows that systems implemented as simple rules that give rise to complex behaviour enable the system itself to ``respond'' to questions about where, in the chain of digits that conform to its behaviour (of a specific node), a certain pattern is to be found. And the fractal nature of information distribution in behaviour allows us to answer complex distribution questions in short forms.

In this section we show the advantages of using an (automatic) perturbation test based on simulation of behaviour using simple rules over the original version of IIT 3.0 based on the ``bottleneck principle''
\cite{oizumi2014fromthe} in computing integrated information.

Taking up the original perturbation test, questions take the form: \emph{what is the output (answer) given this query (input)?.} But in $\phi_{K}$, since questions look for explanations of the behaviour of the system itself, they take the form: \emph{tell me if this pattern is reachable, and if so, tell me where, in the behavioural map, it is possible to find it}.

An example of how $\phi_{K}$ implementations turn into an automatic interrogator is shown in Table 4, which aims to analyze the system networks shown in Figure 8.

\begin{figure}
\begin{centering}
\includegraphics[scale=0.5]{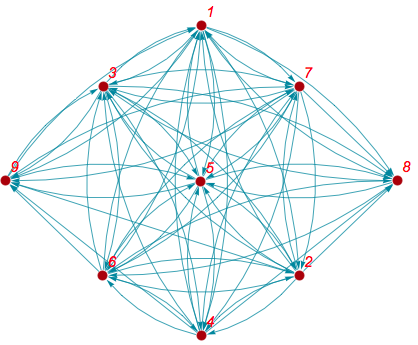} 
\par\end{centering}
\caption{Network example with 9 nodes}
\end{figure}

\begin{table}
\begin{lstlisting}
mmu=MemoryInUse[]; 
findPatternInSharedInputs[{1, 3, 5, 6}, "OR", 1, {1, 5, 7, 3}, "AND", 1]//AbsoluteTiming 
MemoryInUse[]-mmu
\end{lstlisting}

\begin{lstlisting}
{{1, 1, 1, 0, 1}, {1, 1, 1, 1, 1}}}
0.000736
2440
\end{lstlisting}

\caption{$\phi_{K}$ algorithmic querying of the system about its own  behaviour as shown in Figure 8. \textbf{Line 2}: Query: Is it possible for this system to compute \emph{\{8,9\} = \{1,1\} when \{8,9\}-> \{"OR'', "AND''\} and whose input nodes are \{1,3,5,6\} and \{1,5,7,3\}
respectively?. } The results show that the system does compute it
when \{1,3,5,6,7\} = \{\{1,1,1,0,1\},\{1,1,1,1,1\}\}}
\end{table}

In line 2, in the list code shown in Table 4, it is possible to see how $\phi_{K}$ asks questions of a system. This line should be
interpreted as, \emph{Can you compute the pattern \{8,9\} = \{1,1\} when \{8,9\}-> \{"OR'', "AND''\}? If yes, tell me under what conditions you can do so}.

In this example, in the first place $\phi_{K}$ tries to find conditions needed to compute a specific output. As Table 4 shows, the answer is: \emph{Yes! I can. This may happen when \{1,3,5,6,7\} = \{\{1,1,1,0,1\},\{1,1,1,1,1\}\}}. In this answer \{1,3,5,6,7\} is the set of inputs to the subsystem \{8,9\}.

The reader would note here that the answers offered by the system actually are conditions or inputs needed by the system to compute specific input in a format equivalent to Holland's schemas. The schemas' equivalent form for this case would be: \{\{1,{*},1,{*},1,0,1,{*},{*}\},
\{1,{*},1,{*},1,1,1,{*},{*}\}\}, where `{*}' is a wildcard that means 0/1 (any symbol). Such schemas correspond exactly to the generalized answer offered by the system, that is: \{1,3,5,6,7\} = \{\{1,1,1,0,1\},\{1,1,1,1,1\}\}.

This answer, like the Holland's schema theorem \cite{holland1975adaptation}, works by imitating genetics, where a set of genes are responsible for specific features in phenotypes. What $\phi_{K}$ retrieves is the general information that yields specific inputs for the current system.

Probably the greatest advantage of the approach using $\phi_{K}$ in querying samples has to do with the computation time needed to retrieve such information, as compared with a traditional (brute force) approach: 1/10 in this case. (For results of brute force calculation see Appendix, Table 7).

This last suffices as proof that compression and generalization of systems in the form of simple rules based on naturally fractal
information distribution has advantages over common sense or classical approaches to the analysis of complex systems, particularly
in terms of the computational resources needed to compute integrated information.

All the above were applied to analyzing isolated or very simple cases.
In the next section the generalization of $n$ nodes of the system is addressed, and how this works to compute integrated information
according to IIT.

\subsubsection{Shrinking after dividing to rule}

In previous sections it was shown how $\phi_{K}$, applying a perturbation test, can deduce, firstly, what a system is capable of computing and the conditions under which a computation could be performed, and secondly, that by means of simple rules specifying a system it is possible to obtain descriptions of its behaviour in the form of rules that say how information is distributed, or in other words, where, in ordinal terms, such conditions can be found.

The ultimate objective of obtaining this kind of description of the behaviour of a system is to know how many times specific patterns appear in whole repertoires, and thus to construct probability distributions
without need of exorbitant computational resources, since these probability distributions are a key piece used by IIT to compute integrated information.

$\phi_{K}$ addressed such challenges using a two-pronged strategy consisting firstly of parallelizing the analytical process-- which is no more than a technical strategy available to be implemented in almost
any computer language and that falls beyond the scope of this paper--and secondly of the partition of the target sets. This latter part of $\phi_{K}$'s strategy consists of two parts: 1) given a target set to be analyzed, to divide this into parts to be interrogated by $\phi_{K}$ via the implementation of an automatic test, and 2) to find the MIP or the Maximal Information Partition using the algorithm proposed and proved by Oizumi in \cite{kitazono2018efficient}.

In the context of $\phi_{K}$, when a partition of a subject system is being analyzed, the search space for the remaining parts is significantly reduced, facilitating and accelerating the analysis of the remaining parts.

In order to illustrate this idea, take for example Table 5.

\begin{table}
\begin{lstlisting}
cm07 = {
{0, 0, 1, 0, 0, 0, 1},
{0, 0, 1, 0, 0, 1, 0},
{1, 0, 0, 0, 1, 0, 1},
{1, 0, 1, 0, 1, 0, 1},
{0, 0, 1, 1, 0, 1, 1},
{1, 1, 1, 0, 0, 0, 0},
{0, 1, 0, 1, 1, 1, 0}};
dyn07 = {"AND", "OR", "OR", "AND", "OR", "OR", "AND"}; 
onPossibleBehaviour[{1,2,3},{0,0,0},dyn07,cm07]//AbsoluteTiming
onPossibleBehaviour[{2,3,4},{0,0,0},dyn07,cm07]//AbsoluteTiming
onPossibleBehaviour[{1,2,3,4},{0,0,0,0},dyn07,cm07]//AbsoluteTiming
onPossibleBehaviour[{5,6,7},{0,0,0},dyn07,cm07]//AbsoluteTiming
onPossibleBehaviour[{1,2,3,4,5,6,7},{0,0,0,0,0,0,0},dyn07,cm07]//AbsoluteTiming
\end{lstlisting}

\begin{lstlisting}
{0.00076,<|"DecimalRepertoire"->{0},"Sumandos"->{0,2,8,10}|>}
{0.000647,<|"DecimalRepertoire"->{0},"Sumandos"->{0,2,8,10}|>}
{0.0009,<|"DecimalRepertoire"->{0},"Sumandos"->{0,2,8,10}|>}
{0.000656,<|"DecimalRepertoire"->{0,16},"Sumandos"->{0}|>}
{0.001248,<|"DecimalRepertoire"->{0},"Sumandos"->{0}|>}
\end{lstlisting}

\caption{Comparing processing time when a system is divided to compute outputs. \textbf{Line 10-14:} $\phi_{K}$ asking the system defined in lines 1-9 for patterns filled with zeros with different lengths (3, 4 and 7) and combinations. \textbf{Lines 1-5} in results square shown, time in seconds taken for computations and answers in terms of indexes using compressed notation. In first data of results square it
can be observed that the larger the node wanted, the greater the amount of time required to perform the computation, while the time ratio decreases.}
\end{table}

Table 5 shows the definition of a system of 7 nodes (lines 1-9), where a set of a progressively growing length is searched (lines 10-14). In this example $\phi_{K}$  repeatedly asks the system if it is capable of finding a growing pattern of zeros. If it is, the system is requested to show where it is possible to find the desired pattern. Obviously, larger patterns need more computations, but as can be seen in Table 5, in the results square, the time used by $\phi_{K}$ increases as the pattern's length increases (Table 5, results square, lines 1-5), but it grows linearly in contrast to IIT 3.0, where it grows exponentially.

\subsection{Limitations}

When we visualize the behaviour of a system (or subsystem like an isolated node), and take into account its implementation, from the point of view of optimization of computational resources, running rules to generate the behaviour of the whole is still a challenge because it is an expensive process in terms of time and memory. Hence for large systems, analysis based on exhaustive reviews of such behaviour could eventually become intractable.

In order to overcome this limitation, $\phi_{K}$ attempted to find rules that not only give an account of the computability capabilities of a system, but also describe its own behaviour. In other words, we wanted to know about possibilities for finding "shortcuts to express the behaviour'' of a whole system.

One other obvious limitation inherited from computability and algorithmic complexity is that of the semi-computability of the process of trying to find simple representations of behaviour. However, we are not required to find the shortest (simplest) one but simply a set of possible short (simple) ones, which would be an indication of the kind of system we are dealing with. While one can find shorter descriptions using popular lossless compression algorithms, the more powerful the algorithms to find shortcuts and fractal descriptions, the faster the computation and the more telling the results,
something that is to be expected for a relationship between the way in which integrated information is estimated, on the one hand, and algorithmic complexity.

\section{Conclusions and future directions}

\textcolor{black}{Here we have established connections and developed a technique to estimate integrated information using estimations of algorithmic information, which in turn has a solid mathematical foundation. Our computational procedure has targeted what is referred to as the IIT 3.0 version, defined as a calculus of probability distributions. Instead of considering distances between statistical distributions, we formulate the problem as a distance in an algorithmic complexity space, properly approximated, in response to perturbations of the system.}

\textcolor{black}{Interestingly, such a perturbation programmability test--initially inspired by the Turing test (establishing another interesting connection between these new theories of consciousness and old ones)-- as applied to physical systems, is a working strategy to find explanations for the behaviour of systems. It remains for future work to make conceptual and computational connections to what Oizumi and Tononi et al. called the MIP (Minimum Information Partition)~\cite{oizumi2014fromthe} of a system. Having this first version of $\phi_{K}$, we conjecture that MIP definitions also obey and are connected to algorithmic complexity in about the same way, as they should remain based on rules of an algorithmic nature. Thus, the next step is to go further in the application of the test introduced in this paper to discover simple rules that would help to find MIP in a more natural and a faster way. Another possible direction is to systematize the finding of these simple rules and apply more powerful methods to enable computation of larger systems. However, here we have merely established the first principles and the directions that can be explored following these ideas.}

Finally, we think that these ideas about self-explanatory systems capable of providing answers to questions about their own behaviour can help in devising techniques to make other methods, in areas such as machine and deep learning, explain their own, often obscure, behaviour.



\newpage{}

\appendix

\section{Appendix}

\subsection*{How the Meta-perturbation test works}

In order to explain the advantages of the generalization of information
in the form of schemas computed by simple rules, Table 7 is introduced.
In this Table are shown all possible cases where the pattern \{8,9\}=\{1,1\}
can be found in the whole output repertoire of the system introduced
in Figure 8 in the main text.

On the right side of the set contained in Table 7, outputs where \{8,9\}=\{1,1\}
are highlighted in red. On the left side the 9-length are inputs
that yield to outputs containing the desired pattern. On this left
side, in bold, are the corresponding inputs that are particularly
responsible for causing the desired pattern, that is, all possible patterns
for the inputs \{1,3,5,6,7\}

In order to obtain the results in Table 8 using a naive (brute force) approach, it was necessary to define the whole set of all $2^{9}$ possible
inputs and compute the whole set of outputs; then an exhaustive search
for \{8,9\}=\{1,1\} was carried out. Notice that time and memory used
are at least 10 times greater than those used in the $\phi_{K}$ approach.
These results are shown in the last two rows in Table 8 and the
results square in Table 4.

\begin{table}[!htb]
\begin{lstlisting}
cm07 = {{0, 0, 1, 0, 0, 0, 1},
{0, 0, 1, 0, 0, 1, 0},
{1, 0, 0, 0, 1, 0, 1},
{1, 0, 1, 0, 1, 0, 1},
{0, 0, 1, 1, 0, 1, 1},
{1, 1, 1, 0, 0, 0, 0},
{0, 1, 0, 1, 1, 1, 0}};
dyn07 = {"AND", "OR", "OR", "AND", "OR", "OR", "AND"};
analysis07=runDynamic[cm07, dyn07]["RepertoireOutputs"]
\end{lstlisting}

\begin{centering}
\includegraphics[scale=0.5]{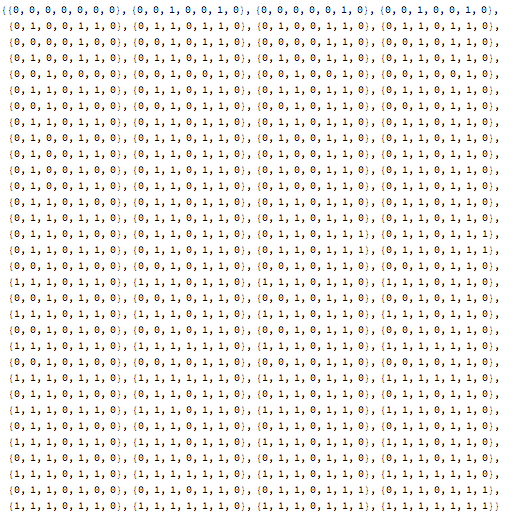} 
\par\end{centering}
\caption{Seven-node system and output repertoires. (A) Network definition
by adjacency matrix (lines 1-7) and dynamics (line 8). (B) Output
repertoire.}
\end{table}

\begin{table}[!htb]
\begin{lstlisting}
{0, 0, 0, 0, 0, 0, 0, 0, 0, 0, 0, 0, 0, 0, 0, 0, 0, 0, 0, 0, 0, 0, 0, 0, 0, 0, 0, 0, 0, 0, 0, 0, 0, 0, 0, 0, 0, 0, 0, 0, 0, 0, 0, 0, 0, 0, 0, 0, 0, 0, 0, 0, 0, 0, 0, 0, 0, 0, 0, 0, 0, 0, 0, 0, 0, 0, 0, 0, 0, 0, 0, 0, 0, 0, 0, 0, 0, 0, 0, 0, 0, 0, 0, 0, 0, 1, 0, 1, 0, 0, 0, 0, 0, 1, 0, 1, 0, 0, 0, 0, 0, 0, 0, 0, 0, 0, 0, 0, 0, 0, 0, 0, 0, 0, 0, 0, 0, 1, 0, 1, 0, 0, 0, 0, 0, 1, 0, 1}
\end{lstlisting}

\begin{lstlisting}
{0, 1, 2, 3, 4, 5, 6, 7, 8, 9, 10, 11, 12, 13, 14, 15, 16, 17, 18, 19, 20, 21, 22, 23, 24, 25, 26, 27, 28, 29, 30, 31, 32, 33, 34, 35, 36, 37, 38, 39, 40, 41, 42, 43, 44, 45, 46, 47, 48, 49, 50, 51, 52, 53, 54, 55, 56, 57, 58, 59, 60, 61, 62, 63, 64, 65, 66, 67, 68, 69, 70, 71, 72, 73, 74, 75, 76, 77, 78, 79, 80, 81, 82, 83, 84, 86, 88, 89, 90, 91, 92, 94, 96, 97, 98, 99, 100, 101, 102, 103, 104, 105, 106, 107, 108, 109, 110, 111, 112, 113, 114, 115, 116, 118, 120, 121, 122, 123, 124, 126}
\end{lstlisting}

\caption{Comparison between real behaviour and unfolded nested rule of behaviour
of node 4 of the system defined in Figure 6 of the main text. The
answer offered by the system (second results square) shows the places
where digit 0 is found in the behaviour of node 4 (first results square)}
\end{table}

\begin{table}[!htb]
\begin{centering}
\{\textbf{1},0,\textbf{1},0,\textbf{1,0,1},0,0\}->\{1,1,0,1,0,0,1,\textbf{\textcolor{black}{1,1}}\}, 
\par\end{centering}
\begin{centering}
\{\textbf{1},1,\textbf{1},0,\textbf{1,0,1},0,0\}->\{1,1,0,1,0,0,1,\textbf{\textcolor{black}{1,1}}\}, 
\par\end{centering}
\begin{centering}
\{\textbf{1},0,\textbf{1},1,\textbf{1,0,1},0,0\}->\{1,1,0,1,0,0,1,\textbf{\textcolor{black}{1,1}}\}, 
\par\end{centering}
\begin{centering}
\{\textbf{1},1,\textbf{1},1,\textbf{1,0,1},0,0\}->\{1,1,0,1,0,0,1,\textbf{\textcolor{black}{1,1}}\}, 
\par\end{centering}
\begin{centering}
\{\textbf{1},0,\textbf{1},0,\textbf{1,1,1},0,0\}->\{1,1,0,1,0,0,1,\textbf{\textcolor{black}{1,1}}\}, 
\par\end{centering}
\begin{centering}
\{\textbf{1},1,\textbf{1},0,\textbf{1,1,1},0,0\}->\{1,1,0,1,0,0,1,\textbf{\textcolor{black}{1,1}}\}, 
\par\end{centering}
\begin{centering}
\{\textbf{1},0,\textbf{1},1,\textbf{1,1,1},0,0\}->\{1,1,0,1,0,0,1,\textbf{\textcolor{black}{1,1}}\}, 
\par\end{centering}
\begin{centering}
\{\textbf{1},1,\textbf{1},1,\textbf{1,1,1},0,0\}->\{1,1,0,1,0,0,1,\textbf{\textcolor{black}{1,1}}\}, 
\par\end{centering}
\begin{centering}
\{\textbf{1},0,\textbf{1},0,\textbf{1,0,1},1,0\}->\{1,1,0,1,0,0,1,\textbf{\textcolor{black}{1,1}}\}, 
\par\end{centering}
\begin{centering}
\{\textbf{1},1,\textbf{1},0,\textbf{1,0,1},1,0\}->\{1,1,0,1,0,0,1,\textbf{\textcolor{black}{1,1}}\}, 
\par\end{centering}
\begin{centering}
\{\textbf{1},0,\textbf{1},1,\textbf{1,0,1},1,0\}->\{1,1,0,1,0,0,1,\textbf{\textcolor{black}{1,1}}\}, 
\par\end{centering}
\begin{centering}
\{\textbf{1},1,\textbf{1},1,\textbf{1,0,1},1,0\}->\{1,1,0,1,0,0,1,\textbf{\textcolor{black}{1,1}}\}, 
\par\end{centering}
\begin{centering}
\{\textbf{1},0,\textbf{1},0,\textbf{1,1,1},1,0\}->\{1,1,0,1,0,0,1,\textbf{\textcolor{black}{1,1}}\}, 
\par\end{centering}
\begin{centering}
\{\textbf{1},1,\textbf{1},0,\textbf{1,1,1},1,0\}->\{1,1,0,1,0,0,1,\textbf{\textcolor{black}{1,1}}\}, 
\par\end{centering}
\begin{centering}
\{\textbf{1},0,\textbf{1},1,\textbf{1,1,1},1,0\}->\{1,1,0,1,0,0,1,\textbf{\textcolor{black}{1,1}}\}, 
\par\end{centering}
\begin{centering}
\{\textbf{1},1,\textbf{1},1,\textbf{1,1,1},1,0\}->\{1,1,0,1,0,0,1,\textbf{\textcolor{black}{1,1}}\}, 
\par\end{centering}
\begin{centering}
\{\textbf{1},0,\textbf{1},0,\textbf{1,0,1},0,1\}->\{1,1,0,1,0,0,1,\textbf{\textcolor{black}{1,1}}\}, 
\par\end{centering}
\begin{centering}
\{\textbf{1},1,\textbf{1},0,\textbf{1,0,1},0,1\}->\{1,1,0,1,0,0,1,\textbf{\textcolor{black}{1,1}}\}, 
\par\end{centering}
\begin{centering}
\{\textbf{1},0,\textbf{1},1,\textbf{1,0,1},0,1\}->\{1,1,0,1,0,0,1,\textbf{\textcolor{black}{1,1}}\}, 
\par\end{centering}
\begin{centering}
\{\textbf{1},1,\textbf{1},1,\textbf{1,0,1},0,1\}->\{1,1,0,1,0,0,1,\textbf{\textcolor{black}{1,1}}\}, 
\par\end{centering}
\begin{centering}
\{\textbf{1},0,\textbf{1},0,\textbf{1,1,1},0,1\}->\{1,1,0,1,0,0,1,\textbf{\textcolor{black}{1,1}}\}, 
\par\end{centering}
\begin{centering}
\{\textbf{1},1,\textbf{1},0,\textbf{1,1,1},0,1\}->\{1,1,0,1,0,0,1,\textbf{\textcolor{black}{1,1}}\}, 
\par\end{centering}
\begin{centering}
\{\textbf{1},0,\textbf{1},1,\textbf{1,1,1},0,1\}->\{1,1,0,1,0,0,1,\textbf{\textcolor{black}{1,1}}\}, 
\par\end{centering}
\begin{centering}
\{\textbf{1},1,\textbf{1},1,\textbf{1,1,1},0,1\}->\{1,1,0,1,0,0,1,\textbf{\textcolor{black}{1,1}}\}, 
\par\end{centering}
\begin{centering}
\{\textbf{1},0,\textbf{1},0,\textbf{1,0,1},1,1\}->\{1,1,0,1,0,0,1,\textbf{\textcolor{black}{1,1}}\}, 
\par\end{centering}
\begin{centering}
\{\textbf{1},1,\textbf{1},0,\textbf{1,0,1},1,1\}->\{1,1,0,1,0,0,1,\textbf{\textcolor{black}{1,1}}\}, 
\par\end{centering}
\begin{centering}
\{\textbf{1},0,\textbf{1},1,\textbf{1,0,1},1,1\}->\{1,1,0,1,0,0,1,\textbf{\textcolor{black}{1,1}}\}, 
\par\end{centering}
\begin{centering}
\{\textbf{1},1,\textbf{1},1,\textbf{1,0,1},1,1\}->\{1,1,0,1,0,1,1,\textbf{\textcolor{black}{1,1}}\}, 
\par\end{centering}
\begin{centering}
\{\textbf{1},0,\textbf{1},0,\textbf{1,1,1},1,1\}->\{1,1,0,1,0,0,1,\textbf{\textcolor{black}{1,1}}\}, 
\par\end{centering}
\begin{centering}
\{\textbf{1},1,\textbf{1},0,\textbf{1,1,1},1,1\}->\{1,1,0,0,0,0,1,\textbf{\textcolor{black}{1,1}}\}, 
\par\end{centering}
\begin{centering}
\{\textbf{1},0,\textbf{1},1,\textbf{1,1,1},1,1\}->\{1,0,0,1,0,0,1,\textbf{\textcolor{black}{1,1}}\}, 
\par\end{centering}
\begin{centering}
\{\textbf{1},1,\textbf{1},1,\textbf{1,1,1},1,1\}->\{1,0,1,0,1,1,1,\textbf{\textcolor{black}{1,1}}\}, 
\par\end{centering}
\begin{centering}
0.166334 
\par\end{centering}
\begin{centering}
217920 
\par\end{centering}
\caption{Naive approach to looking for algorithmic patterns based on simplicity
vs complexity in the calculation of integrated information.}
\end{table}

\section{Appendix}

\IncMargin{1em}
\begin{algorithm}[!htb]
	\SetKwData{Pud}{pud}
	\SetKwData{Fud}{fud}
	\SetKwData{Fbupd}{fbupd}
	\SetKwData{Distros}{distros}
	\SetKwData{ConceptualSpace}{conceptualSpace}
	\SetKwData{Nodes}{nodes}
	\SetKwData{MechaSet}{mechaSet}
	\SetKwData{OneConcept}{OneConcept}
	\SetKwData{BipartitionsSet}{bipartitionsSet}
	\SetKwData{IntegratedInformationValue}{integratedInformationValue}
	\SetKwData{Aux}{aux}
	\SetKwData{UPPD}{UPPD}
	\SetKwData{UFPD}{UFPD}

	\SetKwFunction{computeDistros}{computeDistros}
	\SetKwFunction{Subsets}{Subsets}
	\SetKwFunction{ComputeInputBitProbabilityDistro}{ComputeInputBitProbabilityDistro}
	\SetKwFunction{ComputeOutputBitProbabilityDistro}{ComputeOutputBitProbabilityDistro}
	\SetKwFunction{GetNodes}{GetNodes}
	\SetKwFunction{ComputesPastProbabilityDistribution}{ComputesPastProbabilityDistribution}
	\SetKwFunction{ComputesFutureProbabilityDistribution}{ComputesFutureProbabilityDistribution}
	\SetKwFunction{ComputeConceptOfAMechanism}{ComputeConceptOfAMechanism}
	\SetKwFunction{Append}{Append}
	\SetKwFunction{ComputesConceptualSpace}{ComputesConceptualSpace}
	\SetKwFunction{Bipartitions}{Bipartitions}
	\SetKwFunction{EMD}{EMD}
	\SetKwFunction{GetNodes}{GetNodes}

	\SetKwInOut{Input}{input}
	\SetKwInOut{Output}{output}
	
	\Input{AdjacencyMatrix,Dynamic,CurrentState}
	\Output{Information Integration Value}
	\BlankLine
	\Nodes $\leftarrow$ \GetNodes{AdjacencyMatrix}\;
	\tcp*[h]{UPPD: Unrestricted Past Probability Distribution}\;
	\tcp*[h]{UFPD: Unrestricted Future Probability Distribution}\;
	\UPPD $\leftarrow$  \ComputesPastProbabilityDistribution{\Nodes,CurrentState,$\emptyset$,Dynamic,am}\;
	\UFPD $\leftarrow$  \ComputesFutureProbabilityDistribution{\Nodes,CurrentState,$\emptyset$,Dynamic,am}\;
	\tcp*[h]{am=AdjacencyMatrix; cs=CurrentState}\;
	\ConceptualSpace$\leftarrow$ \ComputesConceptualSpace{am,Dynamic,cs,\UPPD,\UFPD}\;
	
	\IntegratedInformationValue$\leftarrow$ 0\;
	\BipartitionsSet$\leftarrow$ \Bipartitions{\ConceptualSpace}\;

	\ForEach{bipartition $b_i \in \BipartitionsSet$}{%
		\Aux$\leftarrow$\EMD{$b_i,\ConceptualSpace$}\;
		\uIf{ \Aux > \IntegratedInformationValue}{%
			\IntegratedInformationValue$\leftarrow$ \Aux\;
		}
	}

	\caption{computeIntegratedInformation}\label{algo_computeIntegratedInformation}
\end{algorithm}
\DecMargin{1em}

\IncMargin{1em}
\begin{algorithm}[!htb]
	\SetKwData{Pud}{pud}
	\SetKwData{Fud}{fud}
	\SetKwData{Fbupd}{fbupd}
	\SetKwData{Distros}{distros}
	\SetKwData{ConceptualSpace}{conceptualSpace}
	\SetKwData{Nodes}{nodes}
	\SetKwData{MechaSet}{mechaSet}
	\SetKwData{OneConcept}{OneConcept}

	\SetKwFunction{computeDistros}{computeDistros}
	\SetKwFunction{Subsets}{Subsets}
	\SetKwFunction{ComputeInputBitProbabilityDistro}{ComputeInputBitProbabilityDistro}
	\SetKwFunction{ComputeOutputBitProbabilityDistro}{ComputeOutputBitProbabilityDistro}
	\SetKwFunction{GetNodes}{GetNodes}
	\SetKwFunction{ComputesPastProbabilityDistribution}{ComputesPastProbabilityDistribution}
	\SetKwFunction{ComputesFutureProbabilityDistribution}{ComputesFutureProbabilityDistribution}
	\SetKwFunction{ComputeConceptOfAMechanism}{ComputeConceptOfAMechanism}
	\SetKwFunction{Append}{Append}

	\SetKwInOut{Input}{input}
	\SetKwInOut{Output}{output}
	
	\Input{AdjacencyMatrix,Dynamic,CurrentState,UPPD,UFPD}
	\Output{conceptualStructure}
	\BlankLine
	\tcp*[h]{UPPD: Unrestricted Past Probability Distribution}\;
	\tcp*[h]{UFPD: Unrestricted Future Probability Distribution}\;
	\Nodes$\leftarrow$ \GetNodes{AdjacencyMatrix}\;
	\MechaSet$\leftarrow$\Subsets{\Nodes}\;
	
	\ForEach{mechanism $mecha_i \in \MechaSet$}{%
		\OneConcept$\leftarrow$\ComputeConceptOfAMechanism{$mecha_i,\Nodes,CurrentState,UPPD,UFPD$}\;
		\Append{\ConceptualSpace,\OneConcept}
	}

	\caption{computeConceptualSpace}\label{algo_computeConceptualStructure}
\end{algorithm}
\DecMargin{1em}

\IncMargin{1em}
\begin{algorithm}[!htb]
	
	\SetKwData{PurviewsSet}{purviewsSet}
	\SetKwData{APurview}{aPurview}
	\SetKwData{APurviewMIP}{aPurviewMIP}
	\SetKwData{Distros}{distros}
	\SetKwData{Cs}{cs}
	\SetKwData{Connected}{connected}
	\SetKwData{MIP}{MIP}
	\SetKwData{SmallAlpha}{smallAlpha}
	\SetKwData{ConceptualInfo}{ConceptualInfo}
	\SetKwData{PastDistribution}{pastDistribution}
	\SetKwData{FutureDistribution}{futureDistribution}
	
	\SetKwFunction{Subsets}{Subsets}
	\SetKwFunction{Length}{Length}
	\SetKwFunction{Extract}{Extract}
	\SetKwFunction{Part}{Part}
	\SetKwFunction{FullyConnectedQ}{FullyConnectedQ}
	\SetKwFunction{ComputesMIP}{ComputesMIP}
	\SetKwFunction{ComputesDistros}{ComputesDistros}
	\SetKwFunction{EMD}{EMD}
	\SetKwFunction{ComputesPastProbabilityDistribution}{ComputesPastProbabilityDistribution}
	\SetKwFunction{ComputesFutureProbabilityDistribution}{ComputesFutureProbabilityDistribution}
	
	\SetKwInOut{Input}{input}
	\SetKwInOut{Output}{output}
	
	\Input{mechanism,nodesForPurviews,currentState,pastDistro,futDistro,Dynamic, AdjacencyMatrix}
	\Output{concept for current mechanism}
	\BlankLine
	
	\tcp*[h]{nodes where all purviews will be taken from}
	
	\PurviewsSet$\leftarrow$\Subsets{$nodesForPurviews$}\;
	\BlankLine
	
	\For{$j\leftarrow 1$ \KwTo \Length{$PurviewsSet$}}
	{\label{forins}
		
		\APurview$\leftarrow$ \Part{$j,PurviewsSet$}\;
		\Connected$\leftarrow$ \FullyConnectedQ{$mechanism,APurview$}\;
		\lIf{\Connected}{
			
			\tcp*[h]{MIP: Maximal Information Partition}
			
			\SmallAlpha$\leftarrow$ \ComputesMIP{$mechanism,APurview$}
		}
	}
	
	\tcp*[h]{APurviewMIP: Purview responsable to cause MIP for current mechanism}\;
	
	\APurviewMIP$\leftarrow$ \SmallAlpha{$"PurviewMIP"$}\;
	\tcp*[h]{Following sum is formalized in Figure 4, In Text S2 from Oizumi(2014)}\;
	\tcp*[h]{cs=CurrentState; am = AdjacencyMatrix}\;
	\PastDistribution$\leftarrow$ \ComputesPastProbabilityDistribution{$mechanism,cs,\APurviewMIP,\newline
	Dynamic,am$}\;
	\FutureDistribution$\leftarrow$ \ComputesFutureProbabilityDistribution{$mechanism,cs,\APurviewMIP,\newline
	Dynamic,am$}\;
	\ConceptualInfo$\leftarrow$ \EMD{$pastDistro,\PastDistribution$}+\EMD{$futDistro,\FutureDistribution$}\;

	\caption{computeConceptOfAMechanism}\label{algo_computeConceptOfMecha}
\end{algorithm}
\DecMargin{1em}

\IncMargin{1em}
\begin{algorithm}[!htb]
	
	\SetKwData{PurviewsSet}{purviewsSet}	
	\SetKwData{APurview}{aPurview}
	\SetKwData{APurviewMIP}{aPurviewMIP}
	\SetKwData{Distros}{distros}
	\SetKwData{Cs}{cs}
	\SetKwData{Connected}{connected}
	\SetKwData{MIP}{MIP}
	\SetKwData{SmallAlpha}{smallAlpha}
	\SetKwData{ConceptualInfo}{ConceptualInfo}
	\SetKwData{MechaChildren}{mechaChildren}
	\SetKwData{PurviewChildren}{PurviewChildren}
	\SetKwData{Ci}{ci}
	\SetKwData{Ei}{ei}
	\SetKwData{Cei}{cei}
	\SetKwData{PastMIP}{pastMIP}
	\SetKwData{FutMIP}{futMIP}
	\SetKwData{PastDistribution}{pastDistribution}
	\SetKwData{FutureDistribution}{futureDistribution}
	
	\SetKwFunction{Subsets}{Subsets}
	\SetKwFunction{Length}{Length}
	\SetKwFunction{Extract}{Extract}
	\SetKwFunction{Part}{Part}
	\SetKwFunction{FullyConnectedQ}{FullyConnectedQ}
	\SetKwFunction{ComputesMIP}{ComputesMIP}
	\SetKwFunction{ComputesDistros}{ComputesDistros}
	\SetKwFunction{EMD}{EMD}
	\SetKwFunction{ComputesCEI}{ComputesCEI}
	\SetKwFunction{ComputesPastProbabilityDistribution}{ComputesPastProbabilityDistribution}
	\SetKwFunction{ComputesFutureProbabilityDistribution}{ComputesFutureProbabilityDistribution}

	\SetKwInOut{Input}{input}
	\SetKwInOut{Output}{output}
	
	\Input{mechanism,purview}
	\Output{MIP structure}
	\BlankLine
	

	\MechaChildren$\leftarrow$ \Subsets{mechanism}\;
	\PurviewChildren$\leftarrow$ \Subsets{purview}\;
	
	\Ci $\leftarrow$ 10000\;
	\Ei $\leftarrow$ 10000\;

	\ForEach{mecha  $m_i \in \MechaChildren$}{%
		\ForEach{purview $p_i \in \PurviewChildren$}{%
			\tcp*[h]{cs=CurrentState, am=AdjacencyMatrix}\;
			\PastDistribution $\leftarrow$ \ComputesPastProbabilityDistribution{$m_i$,cs,$p_i$,Dynamic,am}\;
			\FutureDistribution $\leftarrow$ \ComputesFutureProbabilityDistribution{$m_i$,cs,$p_i$,Dynamic,am}\;
			\Cei $\leftarrow$ \ComputesCEI{mecha,purview,\PastDistribution,\FutureDistribution}\;
			
			\If{\Cei("ci") < \Ci}{%
				\Ci $\leftarrow$ \Cei("ci")\;
				\PastMIP $\leftarrow$ (mecha,purview)\;
			}
		
			\If{\Cei("ei") < \Ei}{%
				\Ei $\leftarrow$ \Cei("ei")\;
				\FutMIP $\leftarrow$ (mecha,purview)\;
			}

		}
	}
	
\caption{ComputesMIP}\label{algo_computesMIP}
\end{algorithm}
\DecMargin{1em}

\IncMargin{1em}
\begin{algorithm}[!htb]
	
	\SetKwData{PurviewsSet}{purviewsSet}	
	\SetKwData{APurview}{aPurview}
	\SetKwData{APurviewMIP}{aPurviewMIP}
	\SetKwData{Distros}{distros}
	\SetKwData{Cs}{cs}
	\SetKwData{Connected}{connected}
	\SetKwData{MIP}{MIP}
	\SetKwData{SmallAlpha}{smallAlpha}
	\SetKwData{ConceptualInfo}{ConceptualInfo}
	\SetKwData{MechaChildren}{mechaChildren}
	\SetKwData{PurviewChildren}{PurviewChildren}
	\SetKwData{Ci}{ci}
	\SetKwData{Ei}{ei}
	\SetKwData{Cei}{cei}
	\SetKwData{PastMIP}{pastMIP}
	\SetKwData{FutMIP}{futMIP}
	\SetKwData{MechaComplement}{mechaComplement}
	\SetKwData{PurviewComplement}{purviewComplement}
	\SetKwData{PastDistribution}{pastDistribution}
	\SetKwData{FutureDistribution}{futureDistribution}
	\SetKwData{PastDistributionComp}{pastDistributionComp}
	\SetKwData{FutureDistributionComp}{futureDistributionComp}
	
	\SetKwFunction{Subsets}{Subsets}
	\SetKwFunction{Length}{Length}
	\SetKwFunction{Extract}{Extract}
	\SetKwFunction{Part}{Part}
	\SetKwFunction{FullyConnectedQ}{FullyConnectedQ}
	\SetKwFunction{ComputesMIP}{ComputesMIP}
	\SetKwFunction{ComputesDistros}{ComputesDistros}
	\SetKwFunction{EMD}{EMD}
	\SetKwFunction{ComputesCEI}{ComputesCEI}
	\SetKwFunction{Complement}{ComplementI}
	\SetKwFunction{CPPD}{CPPD}
	\SetKwFunction{CFPD}{CFPD}
	\SetKwFunction{Normalize}{Normalize}
	\SetKwFunction{Min}{Min}

	\SetKwInOut{Input}{input}
	\SetKwInOut{Output}{output}
	

	\Input{ChildMecha, ChildPurview, ParentMecha, ParentPurview, ParentPastDistro, ParentFutDistro, UnconstrainedPastDistro, UnconstrainedFutDistro}
	\Output{Causal and Effect information values}
	\BlankLine
	
	\DontPrintSemicolon
	
	\MechaComplement$\leftarrow$ \Complement{$ChildMecha, ParentMecha$}\;
	\PurviewComplement$\leftarrow$ \Complement{$ChildPurview, ParentPurview$}\;
	
	\MechaChildren$\leftarrow$ \Subsets{$mechanism$}\;
	\PurviewChildren$\leftarrow$ \Subsets{$purview$}\;

	\ForEach{mecha  $m_i \in \MechaChildren$}{%
		\ForEach{purview $p_i \in \PurviewChildren$}{%
			ChildMecha $\leftarrow$  $m_i$\;
			ChildPurview $\leftarrow$  $p_i$\;
			
			\uIf{ChildMecha=$\emptyset$}{%
				\PastDistribution $\leftarrow$ UnconstrainedPastDistro\;
				\FutureDistribution $\leftarrow$ UnconstrainedFutDistro\;
				
				\uElseIf{ChildPurview=$\emptyset$}{%
					\PastDistribution $\leftarrow$ 1\;
					\FutureDistribution $\leftarrow$ 1\;
					
					\uElseIf{}{%
						\PastDistribution $\leftarrow$ \ComputesPastProbabilityDistribution{ChildMecha,ChildPurview,cm,am}\;
						\FutureDistribution $\leftarrow$ \ComputesFutureProbabilityDistribution{ChildMecha,ChildPurview,cm,am}\;\;
					}
					
				}
				
			}

			\uIf{\MechaComplement=$\emptyset$}{%
				\PastDistributionComp $\leftarrow$ UnconstrainedPastDistro\;
				\FutureDistributionComp $\leftarrow$ UnconstrainedFutDistro\;
				
				\uElseIf{\PurviewComplement=$\emptyset$}{%
					\PastDistributionComp $\leftarrow$ 1\;
					\FutureDistributionComp $\leftarrow$ 1\;
					
					\uElseIf{}{%
						\tcp*[h]{CPPD=ComputesPastProbabilityDistribution }\;
						\tcp*[h]{CFPD=ComputesFutureProbabilityDistribution}\;
						\PastDistributionComp $\leftarrow$ \CPPD{MechaComplement,PurviewComplement,cm,am}\;
						\FutureDistributionComp $\leftarrow$ \CFPD{MechaComplement,PurviewComplement,cm,am}\;
					}
					
				}
				
			}

			\PastDistribution $\leftarrow$ \Normalize{\PastDistribution*\PastDistributionComp}\;
			\FutureDistribution $\leftarrow$ \Normalize{\FutureDistribution*\FutureDistributionComp}\;
			
			\Ci $\leftarrow$ \EMD{ParentPastDistro,\PastDistribution}\;
			\Ei $\leftarrow$ \EMD{ParentFutDistro,\FutureDistribution}\;
			\Cei $\leftarrow$ \Min{\Ci,\Ei}\;
		}
	}
			
	\caption{computesCEI}\label{algo_computesCEI}
\end{algorithm}
\DecMargin{1em}

\IncMargin{1em}
\begin{algorithm}[!htb]
	
	\SetKwData{PurviewsSet}{purviewsSet}	
	\SetKwData{APurview}{aPurview}
	\SetKwData{APurviewMIP}{aPurviewMIP}
	\SetKwData{Distros}{distros}
	\SetKwData{Cs}{cs}
	\SetKwData{Connected}{connected}
	\SetKwData{MIP}{MIP}
	\SetKwData{SmallAlpha}{smallAlpha}
	\SetKwData{ConceptualInfo}{ConceptualInfo}
	\SetKwData{Mechanism}{mechanism}
	\SetKwData{Purview}{purview}
	\SetKwData{MechaChildren}{mechaChildren}
	\SetKwData{PurviewChildren}{PurviewChildren}
	\SetKwData{Intersec}{intersec}
	\SetKwData{JoinedNames}{joinedNames}
	\SetKwData{Ci}{ci}
	\SetKwData{Ei}{ei}
	\SetKwData{Cei}{cei}
	\SetKwData{PastMIP}{pastMIP}
	\SetKwData{FutMIP}{futMIP}
	\SetKwData{MechaComplement}{mechaComplement}
	\SetKwData{PurviewComplement}{purviewComplement}
	\SetKwData{Sumandos}{sumandos}
	\SetKwData{Powers}{powers}
	\SetKwData{Ins}{ins}
	\SetKwData{Repertoire}{repertoire}
	\SetKwData{Aux}{Aux}
	\SetKwData{Indexes}{indexes}
	
	\SetKwFunction{Subsets}{Subsets}
	\SetKwFunction{Length}{Length}
	\SetKwFunction{Extract}{Extract}
	\SetKwFunction{Part}{Part}
	\SetKwFunction{FullyConnectedQ}{FullyConnectedQ}
	\SetKwFunction{ComputesMIP}{ComputesMIP}
	\SetKwFunction{ComputesDistros}{ComputesDistros}
	\SetKwFunction{EMD}{EMD}
	\SetKwFunction{ComputesCEI}{ComputesCEI}
	\SetKwFunction{Complement}{Complement}	
	\SetKwFunction{Range}{Range}
	\SetKwFunction{Append}{Append}	
	\SetKwFunction{Join}{Join}
	\SetKwFunction{Inputs}{Inputs}
	\SetKwFunction{FirstNode}{FirstNode}
	\SetKwFunction{RepertoireByOutput}{RepertoireByOutput}
	\SetKwFunction{Intersection}{Intersection}
	\SetKwFunction{CreateRepertoire}{CreateRepertoire}
	\SetKwFunction{Combine}{Combine}
	\SetKwFunction{FilterRepertoireByOutput}{FilterRepertoireByOutput}
	\SetKwFunction{Sum}{Sum}

	\SetKwInOut{Input}{input}
	\SetKwInOut{Output}{output}

	\Input{Mechanism,Purview,AdjacencyMatrix,CurrentState,Dynamic}
	\Output{positions (indexes) where current state of mechanism is found into the output repertoire}
	\BlankLine
	
	\DontPrintSemicolon
	
	\tcp*[h]{work as nodes that send inputs to the mechanism. }\;
	\tcp*[h]{This remaning nodes are actually powers that define a pattern }\;
	\tcp*[h]{of distribution of the wanted pattern defined by mechanism}\;
    \BlankLine
	\JoinedNames$\leftarrow$ \Join{\Inputs{$Mechanism$}}\;
	\Powers$\leftarrow$ \Complement{\Range{\Length{AdjacencyMatrix}}, \JoinedNames}-1\;
	\ForEach{node $n_i \in \Powers$}{%
		$\Append(\Sumandos,2^{n_i})$\;
	}
	\Sumandos$\leftarrow$\Subsets{\Sumandos}\;
	\ForEach{sumando $s_i \in \Sumandos$}{%
		$\Append(\Aux,\Sum(s_i))$\;
	}
	\Sumandos$\leftarrow$\Aux\;
	
	\BlankLine
	\BlankLine
	\BlankLine
	\Ins$\leftarrow$ \Inputs{\FirstNode{Mechanism}}\;
	\tcp*[h]{Given a dynamic it computes all possible inputs of defined size that}\;
	\tcp*[h]{results in a defined output (cs)}\;
	
	\Repertoire$\leftarrow$ \RepertoireByOutput{\Length{\Ins},Dynamic,\Cs}\;

	\ForEach{node $m_i \in (\Mechanism-\FirstNode{\Mechanism})$}{%
		\Intersec$\leftarrow$ \Intersection{\Ins,\Inputs{$m_i$}}\;
		\Ins$\leftarrow$ \Ins+(\Inputs{$m_i$}-\Intersec)\;
		\Repertoire$\leftarrow$\Combine{\Repertoire,CreateRepertoire(\Inputs{$m_i$}-\Intersec)}\;
		\Repertoire$\leftarrow$\FilterRepertoireByOutput{\Repertoire,\Cs}\;
	}

	\ForEach{sumando $s_i \in \Sumandos$}{%
		\ForEach{repert $r_i \in \Repertoire$}{%
			$\Append(\Indexes,\Sum(s_i,r_i))$\;
		}	
	}

	\caption{computesPositionsOfAPatternInOutputs}\label{algo_computesPositionsOfAPatternInOutputs}
\end{algorithm}
\DecMargin{1em}

\IncMargin{1em}
\begin{algorithm}[!htb]
	
	\SetKwData{PurviewsSet}{purviewsSet}	
	\SetKwData{APurview}{aPurview}
	\SetKwData{APurviewMIP}{aPurviewMIP}
	\SetKwData{Distros}{distros}
	\SetKwData{Cs}{cs}
	\SetKwData{Connected}{connected}
	\SetKwData{MIP}{MIP}
	\SetKwData{SmallAlpha}{smallAlpha}
	\SetKwData{ConceptualInfo}{ConceptualInfo}
	\SetKwData{Mechanism}{mechanism}
	\SetKwData{Purview}{purview}
	\SetKwData{MechaChildren}{mechaChildren}
	\SetKwData{PurviewChildren}{PurviewChildren}
	\SetKwData{Intersec}{intersec}
	\SetKwData{JoinedNames}{joinedNames}
	\SetKwData{Ci}{ci}
	\SetKwData{Ei}{ei}
	\SetKwData{Cei}{cei}
	\SetKwData{PastMIP}{pastMIP}
	\SetKwData{FutMIP}{futMIP}
	\SetKwData{MechaComplement}{mechaComplement}
	\SetKwData{PurviewComplement}{purviewComplement}
	\SetKwData{Sumandos}{sumandos}
	\SetKwData{Powers}{powers}
	\SetKwData{Ins}{ins}
	\SetKwData{Repertoire}{repertoire}
	\SetKwData{Aux}{Aux}
	\SetKwData{Indexes}{indexes}
	\SetKwData{Locations}{locations}
	\SetKwData{CorrectedLocations}{correctedLocations}
	\SetKwData{Probability}{probability}
	\SetKwData{AllInputs}{allInputs}
	\SetKwData{ProbabilityDistribution}{probabilityDistribution}
	
	\SetKwFunction{Subsets}{Subsets}
	\SetKwFunction{Length}{Length}
	\SetKwFunction{Extract}{Extract}
	\SetKwFunction{Part}{Part}
	\SetKwFunction{FullyConnectedQ}{FullyConnectedQ}
	\SetKwFunction{ComputesMIP}{ComputesMIP}
	\SetKwFunction{ComputesDistros}{ComputesDistros}
	\SetKwFunction{EMD}{EMD}
	\SetKwFunction{ComputesCEI}{ComputesCEI}
	\SetKwFunction{Complement}{Complement}	
	\SetKwFunction{Range}{Range}
	\SetKwFunction{Append}{Append}	
	\SetKwFunction{Join}{Join}
	\SetKwFunction{Inputs}{Inputs}
	\SetKwFunction{FirstNode}{FirstNode}
	\SetKwFunction{RepertoireByOutput}{RepertoireByOutput}
	\SetKwFunction{Intersection}{Intersection}
	\SetKwFunction{CreateRepertoire}{CreateRepertoire}
	\SetKwFunction{Combine}{Combine}
	\SetKwFunction{FilterRepertoireByOutput}{FilterRepertoireByOutput}
	\SetKwFunction{Sum}{Sum}
	\SetKwFunction{ComputesPositionsOfAPattern}{computesPositionsOfAPattern}
	\SetKwFunction{FindPatternInInputs}{FindPatternInInputs}
	\SetKwFunction{Extract}{Extract}
	\SetKwFunction{ComputesProbabilityForElements}{ComputesProbabilityForElements}

	\SetKwInOut{Input}{input}
	\SetKwInOut{Output}{output}


	\Input{Mechanism,Purview,AdjacencyMatrix,CurrentState,Dynamic}
	\Output{Probability distribution for a mechanism}
	\BlankLine
	
	\DontPrintSemicolon
	
	\BlankLine
	\Locations$\leftarrow$\ComputesPositionsOfAPattern{$Mechanism,CurrentState,Purview,AdjacencyMatrix$}\;
	\AllInputs $\leftarrow$ \Extract{\Locations,Purview}\;
	\Probability$\leftarrow$ 1/(\Length{Locations})\;
	
	\ForEach{input $in_i \in \AllInputs$}{%
		\CorrectedLocations$\leftarrow$\FindPatternInInputs{$Purview,in_i,\Length{AdjacencyMatrix}$}
	}
	
	\ProbabilityDistribution $\leftarrow$ \ComputesProbabilityForElements{\CorrectedLocations}

	\caption{computesPastProbabilityDistribution}\label{algo_computesPastProbDistribution}
\end{algorithm}
\DecMargin{1em}

\IncMargin{1em}
\begin{algorithm}[!htb]
	
	\SetKwData{PurviewsSet}{purviewsSet}	
	\SetKwData{APurview}{aPurview}
	\SetKwData{APurviewMIP}{aPurviewMIP}
	\SetKwData{Distros}{distros}
	\SetKwData{Cs}{cs}
	\SetKwData{Connected}{connected}
	\SetKwData{MIP}{MIP}
	\SetKwData{SmallAlpha}{smallAlpha}
	\SetKwData{ConceptualInfo}{ConceptualInfo}
	\SetKwData{Mechanism}{mechanism}
	\SetKwData{Purview}{purview}
	\SetKwData{MechaChildren}{mechaChildren}
	\SetKwData{PurviewChildren}{PurviewChildren}
	\SetKwData{Intersec}{intersec}
	\SetKwData{JoinedNames}{joinedNames}
	\SetKwData{Ci}{ci}
	\SetKwData{Ei}{ei}
	\SetKwData{Cei}{cei}
	\SetKwData{PastMIP}{pastMIP}
	\SetKwData{FutMIP}{futMIP}
	\SetKwData{MechaComplement}{mechaComplement}
	\SetKwData{PurviewComplement}{purviewComplement}
	\SetKwData{Sumandos}{sumandos}
	\SetKwData{Powers}{powers}
	\SetKwData{Ins}{ins}
	\SetKwData{Repertoire}{repertoire}
	\SetKwData{Aux}{Aux}
	\SetKwData{Indexes}{indexes}
	\SetKwData{Locations}{locations}
	\SetKwData{CorrectedLocations}{correctedLocations}
	\SetKwData{Probability}{probability}
	\SetKwData{AllOutputs}{allOutputs}
	\SetKwData{AllInputs}{allInputs}
	\SetKwData{ProbabilityDistribution}{probabilityDistribution}
	
	\SetKwFunction{Subsets}{Subsets}
	\SetKwFunction{Length}{Length}
	\SetKwFunction{Extract}{Extract}
	\SetKwFunction{Part}{Part}
	\SetKwFunction{FullyConnectedQ}{FullyConnectedQ}
	\SetKwFunction{ComputesMIP}{ComputesMIP}
	\SetKwFunction{ComputesDistros}{ComputesDistros}
	\SetKwFunction{EMD}{EMD}
	\SetKwFunction{ComputesCEI}{ComputesCEI}
	\SetKwFunction{Complement}{Complement}	
	\SetKwFunction{Range}{Range}
	\SetKwFunction{Append}{Append}	
	\SetKwFunction{Join}{Join}
	\SetKwFunction{Inputs}{Inputs}
	\SetKwFunction{FirstNode}{FirstNode}
	\SetKwFunction{RepertoireByOutput}{RepertoireByOutput}
	\SetKwFunction{Intersection}{Intersection}
	\SetKwFunction{CreateRepertoire}{CreateRepertoire}
	\SetKwFunction{Combine}{Combine}
	\SetKwFunction{FilterRepertoireByOutput}{FilterRepertoireByOutput}
	\SetKwFunction{Sum}{Sum}
	\SetKwFunction{ComputesPositionsOfAPattern}{computesPositionsOfAPattern}
	\SetKwFunction{FindPatternInInputs}{FindPatternInInputs}
	\SetKwFunction{Extract}{Extract}
	\SetKwFunction{ComputesProbabilityForElements}{ComputesProbabilityForElements}
	\SetKwFunction{ComputesOutputs}{ComputesOutputs}

	\SetKwInOut{Input}{input}
	\SetKwInOut{Output}{output}

	
	\Input{Mechanism,Purview,AdjacencyMatrix,CurrentState,Dynamic}
	\Output{Probability distribution for a mechanism}
	\BlankLine
	
	\DontPrintSemicolon
	
	\BlankLine
	\Locations$\leftarrow$(\FindPatternInInputs{$Purview,CurrentState,\Length{AdjacencyMatrix}$})-1\;
	\AllOutputs $\leftarrow$ \ComputesOutputs{\Locations}\;
	\AllInputs $\leftarrow$ \Extract{\Locations,Purview}\;

	\ProbabilityDistribution $\leftarrow$ \ComputesProbabilityForElements{\AllInputs}

	\caption{computesFutureProbabilityDistribution}\label{algo_computesFutureProbDistribution}
\end{algorithm}
\DecMargin{1em}

\IncMargin{1em}
\begin{algorithm}[!htb]
	
	\SetKwData{PurviewsSet}{purviewsSet}	
	\SetKwData{APurview}{aPurview}
	\SetKwData{APurviewMIP}{aPurviewMIP}
	\SetKwData{Distros}{distros}
	\SetKwData{Cs}{cs}
	\SetKwData{Connected}{connected}
	\SetKwData{MIP}{MIP}
	\SetKwData{SmallAlpha}{smallAlpha}
	\SetKwData{ConceptualInfo}{ConceptualInfo}
	\SetKwData{Mechanism}{mechanism}
	\SetKwData{Purview}{purview}
	\SetKwData{MechaChildren}{mechaChildren}
	\SetKwData{PurviewChildren}{PurviewChildren}
	\SetKwData{Intersec}{intersec}
	\SetKwData{JoinedNames}{joinedNames}
	\SetKwData{Ci}{ci}
	\SetKwData{Ei}{ei}
	\SetKwData{Cei}{cei}
	\SetKwData{PastMIP}{pastMIP}
	\SetKwData{FutMIP}{futMIP}
	\SetKwData{MechaComplement}{mechaComplement}
	\SetKwData{PurviewComplement}{purviewComplement}
	\SetKwData{Sumandos}{sumandos}
	\SetKwData{Powers}{powers}
	\SetKwData{Ins}{ins}
	\SetKwData{Repertoire}{repertoire}
	\SetKwData{Aux}{Aux}
	\SetKwData{Indexes}{indexes}
	\SetKwData{Locations}{locations}
	\SetKwData{Probability}{probability}
	\SetKwData{Limit}{limit}
	\SetKwData{Repetitions}{repetitions}
	\SetKwData{Longi}{longi}
	\SetKwData{Serie}{serie}
	\SetKwData{Found}{Found}

	\SetKwFunction{Subsets}{Subsets}
	\SetKwFunction{Length}{Length}
	\SetKwFunction{Extract}{Extract}
	\SetKwFunction{Part}{Part}
	\SetKwFunction{FullyConnectedQ}{FullyConnectedQ}
	\SetKwFunction{ComputesMIP}{ComputesMIP}
	\SetKwFunction{ComputesDistros}{ComputesDistros}
	\SetKwFunction{EMD}{EMD}
	\SetKwFunction{ComputesCEI}{ComputesCEI}
	\SetKwFunction{Complement}{Complement}	
	\SetKwFunction{Range}{Range}
	\SetKwFunction{Append}{Append}	
	\SetKwFunction{Join}{Join}
	\SetKwFunction{Inputs}{Inputs}
	\SetKwFunction{FirstNode}{FirstNode}
	\SetKwFunction{RepertoireByOutput}{RepertoireByOutput}
	\SetKwFunction{Intersection}{Intersection}
	\SetKwFunction{CreateRepertoire}{CreateRepertoire}
	\SetKwFunction{Combine}{Combine}
	\SetKwFunction{FilterRepertoireByOutput}{FilterRepertoireByOutput}
	\SetKwFunction{Sum}{Sum}
	\SetKwFunction{ComputesPositionsOfAPattern}{computesPositionsOfAPattern}
	\SetKwFunction{CreateSerieOddNumbers}{CreateSerieOddNumbers}
	\SetKwFunction{CreateSerieEvenNumbers}{CreateSerieEvenNumbers}

	\SetKwInOut{Input}{input}
	\SetKwInOut{Output}{output}

	
	\Input{Nodes,WantedPattern,sizeAdjacencyMatrix}
	\Output{Finds indexes in input repertoire where nodes fullfill wantedPattern}
	\BlankLine
	
	\DontPrintSemicolon
	
	$\Limit \leftarrow 2^{sizeAdjacencyMatrix}$\;
	
	\ForEach{node $n_i \in Nodes$}{%
			$\Powers \leftarrow 2^{n_i-1}$\;
			$\Repetitions \leftarrow \Limit/\Powers$\;
			$\Longi \leftarrow \Limit/\Repetitions$\;
			
			\uIf{if expectedPatt = 1}{
				$\Serie \leftarrow \CreateSerieEvenNumbers(\Repetitions)$\;
			}
			\Else{
				$\Serie \leftarrow \CreateSerieOddNumbers(\Repetitions)$\;
			}
			
			\For{$i = 1;\ i < \Length{\Serie};\ i = i + 1$}{
				$\Found \leftarrow \Range{((\Powers*\Serie[i])-\Longi)+1, \Powers*\Serie[i]}$\;
			}

	}

	\caption{findPatternInInputs}\label{algo_findPatternsInInputs}
\end{algorithm}
\DecMargin{1em}

\IncMargin{1em}
\begin{algorithm}[!htb]
	
	\SetKwData{PurviewsSet}{purviewsSet}	
	\SetKwData{APurview}{aPurview}
	\SetKwData{APurviewMIP}{aPurviewMIP}
	\SetKwData{Distros}{distros}
	\SetKwData{Cs}{cs}
	\SetKwData{Connected}{connected}
	\SetKwData{MIP}{MIP}
	\SetKwData{SmallAlpha}{smallAlpha}
	\SetKwData{ConceptualInfo}{ConceptualInfo}
	\SetKwData{Mechanism}{mechanism}
	\SetKwData{Purview}{purview}
	\SetKwData{MechaChildren}{mechaChildren}
	\SetKwData{PurviewChildren}{PurviewChildren}
	\SetKwData{Intersec}{intersec}
	\SetKwData{JoinedNames}{joinedNames}
	\SetKwData{Ci}{ci}
	\SetKwData{Ei}{ei}
	\SetKwData{Cei}{cei}
	\SetKwData{PastMIP}{pastMIP}
	\SetKwData{FutMIP}{futMIP}
	\SetKwData{MechaComplement}{mechaComplement}
	\SetKwData{PurviewComplement}{purviewComplement}
	\SetKwData{Sumandos}{sumandos}
	\SetKwData{Powers}{powers}
	\SetKwData{Ins}{ins}
	\SetKwData{Repertoire}{repertoire}
	\SetKwData{Aux}{Aux}
	\SetKwData{Indexes}{indexes}
	\SetKwData{Locations}{locations}
	\SetKwData{Probability}{probability}
	\SetKwData{Limit}{limit}
	\SetKwData{Repetitions}{repetitions}
	\SetKwData{Longi}{longi}
	\SetKwData{Serie}{serie}
	\SetKwData{Found}{Found}
	\SetKwData{ZeroProbability}{zeroProbability}
	\SetKwData{OneProbability}{oneProbability}

	\SetKwFunction{Subsets}{Subsets}
	\SetKwFunction{Length}{Length}
	\SetKwFunction{Extract}{Extract}
	\SetKwFunction{Part}{Part}
	\SetKwFunction{FullyConnectedQ}{FullyConnectedQ}
	\SetKwFunction{ComputesMIP}{ComputesMIP}
	\SetKwFunction{ComputesDistros}{ComputesDistros}
	\SetKwFunction{EMD}{EMD}
	\SetKwFunction{ComputesCEI}{ComputesCEI}
	\SetKwFunction{Complement}{Complement}	
	\SetKwFunction{Range}{Range}
	\SetKwFunction{Append}{Append}	
	\SetKwFunction{Join}{Join}
	\SetKwFunction{Inputs}{Inputs}
	\SetKwFunction{FirstNode}{FirstNode}
	\SetKwFunction{RepertoireByOutput}{RepertoireByOutput}
	\SetKwFunction{Intersection}{Intersection}
	\SetKwFunction{CreateRepertoire}{CreateRepertoire}
	\SetKwFunction{Combine}{Combine}
	\SetKwFunction{FilterRepertoireByOutput}{FilterRepertoireByOutput}
	\SetKwFunction{Sum}{Sum}
	\SetKwFunction{ComputesPositionsOfAPattern}{computesPositionsOfAPattern}
	\SetKwFunction{CreateSerieOddNumbers}{CreateSerieOddNumbers}
	\SetKwFunction{CreateSerieEvenNumbers}{CreateSerieEvenNumbers}
	\SetKwFunction{FindPatternInInputs}{FindPatternInInputs}

	\SetKwInOut{Input}{input}
	\SetKwInOut{Output}{output}

	
	\Input{Nodes,AdjacencyMatrix,Dynamics}
	\Output{Bit probability for given nodes}
	\BlankLine
	
	\DontPrintSemicolon
	
	$\Locations \leftarrow \FindPatternInInputs{Nodes,1,\Length{AdjacencyMatrix}}$\;
	$\OneProbability \leftarrow \Length{\Locations}/2^{\Length{AdjacencyMatrix}}$\;
	$\ZeroProbability \leftarrow 1-\OneProbability$\;

	\caption{computeInputBitProbabilityDistro}\label{algo_computeInputBitProbabilityDistro}
\end{algorithm}
\DecMargin{1em}

\IncMargin{1em}
\begin{algorithm}[!htb]
	
	\SetKwData{PurviewsSet}{purviewsSet}	
	\SetKwData{APurview}{aPurview}
	\SetKwData{APurviewMIP}{aPurviewMIP}
	\SetKwData{Distros}{distros}
	\SetKwData{Cs}{cs}
	\SetKwData{Connected}{connected}
	\SetKwData{MIP}{MIP}
	\SetKwData{SmallAlpha}{smallAlpha}
	\SetKwData{ConceptualInfo}{ConceptualInfo}
	\SetKwData{Mechanism}{mechanism}
	\SetKwData{Purview}{purview}
	\SetKwData{MechaChildren}{mechaChildren}
	\SetKwData{PurviewChildren}{PurviewChildren}
	\SetKwData{Intersec}{intersec}
	\SetKwData{JoinedNames}{joinedNames}
	\SetKwData{Ci}{ci}
	\SetKwData{Ei}{ei}
	\SetKwData{Cei}{cei}
	\SetKwData{PastMIP}{pastMIP}
	\SetKwData{FutMIP}{futMIP}
	\SetKwData{MechaComplement}{mechaComplement}
	\SetKwData{PurviewComplement}{purviewComplement}
	\SetKwData{Sumandos}{sumandos}
	\SetKwData{Powers}{powers}
	\SetKwData{Ins}{ins}
	\SetKwData{Repertoire}{repertoire}
	\SetKwData{Aux}{Aux}
	\SetKwData{Indexes}{indexes}
	\SetKwData{Locations}{locations}
	\SetKwData{Probability}{probability}
	\SetKwData{Limit}{limit}
	\SetKwData{Repetitions}{repetitions}
	\SetKwData{Longi}{longi}
	\SetKwData{Serie}{serie}
	\SetKwData{Found}{Found}
	\SetKwData{ZeroProbability}{zeroProbability}
	\SetKwData{OneProbability}{oneProbability}

	\SetKwFunction{Subsets}{Subsets}
	\SetKwFunction{Length}{Length}
	\SetKwFunction{Extract}{Extract}
	\SetKwFunction{Part}{Part}
	\SetKwFunction{FullyConnectedQ}{FullyConnectedQ}
	\SetKwFunction{ComputesMIP}{ComputesMIP}
	\SetKwFunction{ComputesDistros}{ComputesDistros}
	\SetKwFunction{EMD}{EMD}
	\SetKwFunction{ComputesCEI}{ComputesCEI}
	\SetKwFunction{Complement}{Complement}	
	\SetKwFunction{Range}{Range}
	\SetKwFunction{Append}{Append}	
	\SetKwFunction{Join}{Join}
	\SetKwFunction{Inputs}{Inputs}
	\SetKwFunction{FirstNode}{FirstNode}
	\SetKwFunction{RepertoireByOutput}{RepertoireByOutput}
	\SetKwFunction{Intersection}{Intersection}
	\SetKwFunction{CreateRepertoire}{CreateRepertoire}
	\SetKwFunction{Combine}{Combine}
	\SetKwFunction{FilterRepertoireByOutput}{FilterRepertoireByOutput}
	\SetKwFunction{Sum}{Sum}
	\SetKwFunction{ComputesPositionsOfAPattern}{computesPositionsOfAPattern}
	\SetKwFunction{CreateSerieOddNumbers}{CreateSerieOddNumbers}
	\SetKwFunction{CreateSerieEvenNumbers}{CreateSerieEvenNumbers}
	\SetKwFunction{ComputesPositionsOfAPatternInOutputs}{ComputesPositionsOfAPatternInOutputs}

	\SetKwInOut{Input}{input}
	\SetKwInOut{Output}{output}

	
	\Input{Nodes,AdjacencyMatrix,Dynamics}
	\Output{Bit probability for given nodes}
	\BlankLine
	
	\DontPrintSemicolon
	
	$\Locations \leftarrow \ComputesPositionsOfAPatternInOutputs{Nodes,1,Dynamics,AdjacencyMatrix}$\;
	$\OneProbability \leftarrow \Length{\Locations}/2^{\Length{AdjacencyMatrix}}$\;
	$\ZeroProbability \leftarrow 1-\OneProbability$\;

	\caption{computeOutputBitProbabilityDistro}\label{algo_computeOutputBitProbabilityDistro}
\end{algorithm}
\DecMargin{1em}


\end{document}